\documentclass[12pt,letterpaper]{JHEP3}         
\usepackage{epsfig}

\def\be{\begin{eqnarray}}
\def\ee{\end{eqnarray}}
\newcommand{\nn}{\nonumber}
\newcommand\para{\paragraph{}}
\newcommand{\ft}[2]{{\textstyle\frac{#1}{#2}}}
\newcommand{\tfrac}[2]{{\frac{#1}{#2}}}
\newcommand{\eqn}[1]{(\ref{#1})}

\def\Dslash{\,\,{\raise.15ex\hbox{/}\mkern-12mu D}}
\def\Dbarslash{\,\,{\raise.15ex\hbox{/}\mkern-12mu {\bar D}}}
\def\delslash{\,\,{\raise.15ex\hbox{/}\mkern-9mu \partial}}
\def\delbarslash{\,\,{\raise.15ex\hbox{/}\mkern-9mu {\bar\partial}}}
\def\pslash{\,\,{\raise.15ex\hbox{/}\mkern-9mu p}}
\def\calDslash{\,\,{\raise.15ex\hbox{/}\mkern-12mu {\cal D}}}

\newcommand{\RN}{Reissner-Nordstr\"om}

\def\lae{\mathrel{\mathop{\smash{\lower .5 ex \hbox{$\stackrel<\sim$}}}}}
\def\lae{\mathrel{\mathop{\smash{\lower .5 ex \hbox{$\stackrel>\sim$}}}}}

\title{Holographic Dual of the Lowest Landau Level}

\author{Mike Blake${}^1$, Stefano Bolognesi${}^2$,
David Tong${}^{1}$ and Kenny Wong${}^1$ \\

${}^1$ Department of Applied Mathematics and Theoretical Physics, University of Cambridge, UK \\
${}^2$ Racah Institute of Physics, The Hebrew University of Jerusalem, Israel
\\ {\ } \\
{\tt stefanobolo@gmail.com, m.a.blake, d.tong, k.wong@damtp.cam.ac.uk}
}

\abstract{We describe the lowest Landau level of a quantum electron star in AdS$_4$. In the presence of a suitably strong magnetic field, the dynamics of fermions in the bulk is effectively reduced from four to two dimensions. These two-dimensional fermions can subsequently be treated using  the techniques of bosonization and the difficult many-body problem of building a gravitating, charged quantum star is reduced to solving the sine-Gordon model coupled to a gauge field and a metric.  The kinks of the sine-Gordon model provide the  holographic dual of the lowest Landau levels of the strongly-coupled $d=2+1$-dimensional boundary field theory. The system exhibits order one oscillations in the magnetic  susceptibility, now arising as a classical effect in the bulk.  Moreover, as the chemical potential is varied, we find jumps in the charge density, oscillations in the fractionalised charge density and plateaux in the cohesive charge density.}

\begin{document}
\pagestyle{plain} \setcounter{page}{1}
\newcounter{bean}
\baselineskip16pt

\section{Introduction}

If you want to construct a gravitating star out of quantum fields, it's a lot easier to work with  bosons than  fermions. For bosons, you can place a macroscopic number of particles in the same state and, correspondingly, you need only find  a single solution of the coupled Einstein-scalar equations. For fermions, the exclusion principle means this isn't possible: instead  you need to solve the coupled Einstein-Dirac equations $10^{23}$ times. 

\para
Of course, to describe stars in the night sky, a fortuitous   separation of scales means that you don't need to start with quantum fields. The Compton wavelength of the particles is much (much) less than the curvature of space. This ensures that you can  treat the fermions as a fluid of particles, using standard statistical mechanics techniques to compute their equation of state and stress tensor which subsequently feed into the Einstein equations. However, in situations where the Compton wavelength of the particles is comparable to the curvature of space, you are left with a formidable many-body problem: the construction of a genuinely quantum star. 

\para
The problem of constructing quantum stars has recently arisen in the search for 
holographic models describing a finite density of strongly interacting fermions. Initial attempts to understand  fermions at finite density focussed on probe Dirac fields in an AdS black hole background \cite{sungsik,mit1,leiden,mit2}. This class of models was shown to encompass a large variety of Fermi liquids, non-Fermi liquids and marginal Fermi liquids. See \cite{review} for a detailed review.

\para
While the results of the probe calculations are extremely encouraging, they leave a number of unanswered questions. Most notably,  the charge density in the boundary theory is not carried by the fermions themselves, but is rather contained in `fractionalised' charge, hidden behind the horizon of the black hole.  Moreover, the horizon suffers from a non-zero entropy density at zero temperature, in violation of the third law of thermodynamics and strongly suggesting that the black hole is not the true ground state of the system. Indeed, one important instability was identified in \cite{hpst}: the fermions want to develop a charge density outside the horizon. This charge density is sufficient to cause appreciable  backreaction, both screening the electric field of the black hole and, ultimately, replacing the horizon with a different geometry with vanishing entropy. The resulting configuration is referred to as an {\it electron star} \cite{sean1}. 

\para
The phenomenology of electron stars was subsequently explored in a number of papers \cite{sean2,sean3,lars1,lars2}.  However, to avoid the complexities describe above, all these papers work in a regime in which the mass, $m$, of the fermion is large compared to $L$, the AdS radius: $mL\gg 1$. This provides the necessary separation of scales to model the equation of the state of the star  using a simple fluid description. While this makes the calculations computationally tractable, it is not without consequence. For example, the electron star  exhibits a large number of densely packed Fermi surfaces, corresponding to the radial decomposition of the fermion field in the bulk \cite{sean3,hong}. The existence of many Fermi surfaces is not particularly desirable and can be traced directly to the need to work in $mL\gg 1$ approximation.

\para
More recently, progress has been made in constructing a truly quantum electron star in a regime where $mL\sim 1$ \cite{sachdev,mcgreevy}. If you're only interested in the geometry rather than the fermionic wavefunctions, you don't really need to solve the equations $10^{23}$ times; you only need to calculate the determinant of the Dirac operator in a state which is occupied by $10^{23}$ fermions. However, this is still a technically challenging subject and is not yet as well developed as the earlier approach.

\para
The purpose of this paper is to point out that there is a regime in which the full quantum electron star in AdS$_4$ can be constructed with relative ease. This is the regime of large magnetic field, $B$. By this we mean $B> \mu^2$, where $\mu$ is the chemical potential of the boundary field theory. 
\para
There is a simple, intuitive explanation for why electron stars are particularly simple in this limit. In the presence of a large magnetic field, the bulk electrons are  forced into their lowest Landau level. This results in a kind of dimensional reduction of their dynamics:   the electrons are trapped in the spatial directions parallel to the boundary and only free to move in the radial direction of AdS. In other words, the fermions become effectively two dimensional.

\para
But fermions in two dimensions are special. They are equivalent to bosons. This means that the problem involving a finite density of quantum fermions can be mapped to a classical problem of a scalar field, coupled to a gauge field and metric. And, as we mentioned in the opening paragraph, this is much simpler to solve. Indeed, at heart the problem of constructing a quantum electron star becomes morally equivalent to the problem of screening an electric field by fermions in two dimensions \cite{coleman}: we will find that we need to solve a variant of the sine-Gordon model in AdS-like geometries. This approach also provides us with the analog of the fermion wavefunctions, providing an intuitive visualisation of the distribution of fermionic charge in the bulk.

\para
The boundary dual of our  holographic model is a $2+1$-dimensional, strongly-coupled, field theory. The theory is studied at finite density, with a  chemical potential $\mu$, and in the presence of a magnetic field $B$. We are restricting ourselves to the limit 
\be B > \mu^2
\nn\ee
If the boundary theory consisted of free fermions, this requirement would force us to the lowest Landau level. We will find that something similar holds for our strongly interacting theory. We are now dealing with large $N$ gauge theory and, correspondingly, the gauge invariant fermionic operators have a large number of ``lowest Landau levels", corresponding to the radial harmonics of the fermionic field in the bulk. We will refer to these different Landau levels as different ``bands". 

\para
Our construction of electron stars in the lowest Landau level allows us to exhibit a number of quantum phenomena using purely classical bulk physics. These include discrete jumps in the charge density as the chemical potential is varied and de Haas van Alphen oscillations as the magnetic field is varied.


\para 
 The organisation of the paper is as follows. Section 2 is devoted to the bulk. We study the decomposition of fermions in a magnetic field and explain how one can bosonize those fermions in the lowest Landau level. We further review some well known aspects of bosonization, including the existence of kinks and their relationship to the anomaly, both of which gain a slight twist in the present context.  In Section 3, we solve our bosonized equations to construct quantum electron stars in both hard wall and black hole geometries.  We subsequently  use these solutions to explore various phenomena in the $d=2+1$ dimensional boundary theory, such as  oscillations in the magnetic  susceptibility  and the interplay between filling fractions and the  fractionalisation of charge due to the black hole horizon. A number of technical issues are relegated to appendices.


\section{Electron Stars in the Lowest Landau Level}

Our starting point for constructing magnetised electron stars is the Einstein-Maxwell-Dirac action with negative cosmological constant. 
\be
S= \int d^4x\sqrt{g}\ \left[\frac{1}{2\kappa^2}\left(R + \frac{6}{L^2}\right) -\frac{1}{4e^2}F_{\mu\nu}F^{\mu\nu}  
-i\bar{\psi}\left(e^\mu_a\Gamma^a{\cal D}_\mu - m\right)\psi\right]\label{action}\ee
%
%
%
Here the field strength is $F=dA$, while $\psi$ is a four-component Dirac fermion\footnote{Our conventions: The covariant derivative contains both spin and gauge connections, ${\cal D}_\mu\psi \equiv (\partial_\mu -iA_\mu + \ft18\omega_{\mu,bc}[\Gamma^a,\Gamma^b])\psi$.  The vierbein $e^\mu_a$ translates between tangent space indices, $a,b$ and spacetime indices $\mu,\nu$. Our gamma matrices carry tangent space indices and obey the Clifford algebra Clifford algebra $\{\Gamma^a, \Gamma^b\} = 2\eta^{ab} $. We define the chiral gamma matrix as $\Gamma^5 = - i \Gamma^{t} \Gamma^{r} \Gamma^{x} \Gamma^{y}$. The gamma matrices have the hermiticity property $\Gamma^{t} (\Gamma^a)^\dagger \Gamma^{t}=\Gamma^{a}$ and conjugate spinors are defined by $\bar\psi = \psi^\dagger \Gamma^t$.}. 
Our solutions will be asymptotically AdS$_4$ with radius $L$. We choose radial coordinate, $r$, such that the boundary lies at $r=0$.  

\para
We impose standard boundary conditions on all fields. For the fermions, this means
\be (1-\Gamma^r)\psi=0\ \qquad {\rm at}  \ r=0\label{fermibc}\ee
while the gauge field has Dirichlet boundary conditions. Our  interest lies in dual boundary theories at finite density, with chemical potential $\mu$, and in the presence of a magnetic field, $B$. These are implemented through the requirement that the gauge fields asymptote to 
\be A_t \rightarrow -\mu\ \ \ \ ,\ \ \ \ A_y\rightarrow Bx\qquad {\rm as}\ r\rightarrow 0\label{muB}\ee
Here we have chosen Landau gauge for the magnetic field. As usual, the gauge field is not translationally invariant but all gauge invariant quantities, such as the field strength $F_{xy}$, are invariant under translations.  In this paper, we restrict our attention to such translationally invariant solutions, a seemingly benign fact but one which will have important consequences later. The most general form of the metric is
%
%
\be
ds^2 = \frac{L^2}{r^2} \left( -f(r) dt^2 + h(r) dr^2 + dx^2 + dy^2 \right).
\label{metric}\ee
where $f(r), h(r)\rightarrow 1$ as $r\rightarrow 0$. The temporal gauge field also gains a radial dependence, $A_t=A_t(r)$. However, in the absence of any magnetic monopoles in the bulk, the Maxwell equations require that the magnetic field is constant\footnote{Physics in the presence of bulk magnetic monopoles was studied in \cite{adsmono}.}.

\para
The equations of motion arising from \eqn{action}, subject to the boundary conditions \eqn{muB}, admit a family of well known  solutions given by the dyonic AdS Reissner-Nordstr\"om black hole. 
In this paper we work only at zero temperature. The metric for the extremal Reissner-Nordstr\"om black hole  takes the form \eqn{metric} with 
\be f(r) = \frac{1}{h(r)} = 1 -  \frac{4r^3}{r_h^3} + \frac{3r^4}{r_h^4}\label{rn1}\ee
where the black hole horizon, $r_h$, lies at the root of the quadratic
\be B^2 r_h^4 + \mu^2 r_h^2 - \frac{6e^2L^2}{\kappa^2}=0\label{rn2}\ee
The temporal component of the gauge field has the profile
\be A_t = -\mu\left(1-\frac{r}{r_h}\right)\nn\ee

\para
Importantly, the fermions are not excited in the Reissner-Nordstr\"om background. Yet this need not be their ground state. As first shown in \cite{hpst}, and explored in some detail in \cite{sean1,sean2,sean3}, for electrically charged black holes the local bulk chemical potential may be large enough to excite a bulk fermionic charge density which screens the electric field $F_{tr}$ and backreacts on the geometry. The resulting configuration is the electron star.  The goal of constructing an electron star is to find further solutions $f(r)$, $h(r)$ and $A_t(r)$ supported by fermions.

\para
As explained in the introduction, a full understanding of the electron star has only been achieved in certain regimes of parameter space. This is best developed in the limit $mL\gg 1$, where the Compton wavelength of the fermions is much smaller than the curvature scale of spacetime. Here a Thomas-Fermi approximation can be applied for the density of states and the fermions are subsequently treated as a perfect, charged fluid coupled to gravity in the usual Tolman-Oppenheimer-Volkov manner. Such a method was also applied to study neutral (neutron) stars in AdS \cite{jan1,jan2}.  

\para
Electron stars have also been studied in the presence of a magnetic field. Small fields, $B\ll \mu^2$ --- which is relevant if comparing to, say, strange metal phenomenology -- were studied in \cite{sean2} where magnetic oscillations, including the Kosevich-Lifshitz formula, were recovered as quantum effects in the bulk. More recently, the effect of magnetic fields on electron stars in theories with dilaton couplings were discussed in \cite{clifford}. Both of these works treat the electron stars in the fluid, Thomas-Fermi approximation. 

\para
The purpose of this paper is to show that there is another regime where the problem of electron stars becomes tractable. This is the regime of large magnetic fields
\be B > \mu^2\label{bigger}\ee
Here the physics is dominated by the lowest Landau level, both in the bulk and, in an appropriate sense, in the boundary. Indeed, one could say that focus  of this paper is to understand what it means to be in the lowest Landau level of the strongly interacting boundary theory. 

\para
Let us first recall some simple facts about free, massless Dirac fermions in $d=2+1$ dimensions. 
 With a magnetic field turned on, the spectrum rearranges itself into a  tower of relativistic Landau levels, with energies  
 $E_n = \sqrt{2Bn}$, for $n=0,1,2,\ldots$.  
 \para
  The lowest, $n=0$, Landau level has vanishing energy and is spin polarized: the usual $+B/2$ zero point energy of the lowest Landau level is cancelled by a $-B/2$ Zeeman splitting for the spin down electrons. The net result is that the degeneracy of electrons in a plane of area $A$ is $BA/2\pi$ for the $n=0$ level. All higher Landau levels have degeneracy  $BA/\pi$.  

\para
For free fermions placed in a magnetic field $B>\mu^2/2$, the lowest Landau level is filled, with all others empty.   As we will see in Section \ref{3sec}, the story is much richer for the the strongly interacting $d=2+1$ dimensional theory that lives on the boundary of AdS$_4$. In the remainder of this section, we will show how to construct  a bulk electron star in the regime \eqn{bigger}.  For such large magnetic fields, it is possible to make progress even when the Compton wavelength is comparable to the AdS radius, $mL\sim 1$. In other words, we construct a quantum electron star, albeit restricted to its lowest Landau level.

\subsection{The Landau Levels of the Electron Star}

\para
Our goal is to construct  the bulk electron star in the regime $\mu < \sqrt{B}$. To proceed, we decompose the bulk fermions into Landau levels in the $x-y$ plane, each of which is still free to move in the radial direction. Aspects of fermions in magnetic AdS geometries have been previously discussed in \cite{albash2,albash,elena,magcat,magcat2}. 

\para
With one eye to later bosonization, it will prove useful to first change coordinate system. Although the metric \eqn{metric} is more familiar, the `tortoise' coordinate, $\tilde{r}$, defined such that  the $(\tilde{r}-t)$ plane is conformally flat, is calculationally simpler. We therefore work with the metric
\be
ds^2 = \Omega^2(\tilde{r}) (-dt^2 + d\tilde{r}^2) + \Sigma^2(\tilde{r}) (dx^2 + dy^2)\label{metric2}
\ee
where 
\be \frac{d\tilde{r}}{dr} = \sqrt{\frac {h(r)}{f(r)}} \ \ \ ,\ \ \  \Omega^2(\tilde r) = f(r)\frac {L^2} {r^2}\ \ \ , \ \ \  
\Sigma^2(\tilde r) = \frac {L^2} {r^2}\label{thoseguys}\ee
It is a simple matter to write the action \eqn{action} for the Dirac fermion $\psi$ propagating in this background, together with a gauge potential $A_t(\tilde{r})$ and bulk magnetic field $B$. However, when dealing with fermions in curved spacetime, it is often simplest to perform a conformal transformation, which amounts to a simple rescaling of the fermion 
%
%
%
%
\be \tilde{\psi} = \sqrt{\Omega}\Sigma\,\psi\nn\ee
In these variables, the Dirac action \eqn{action} takes the form
\be
S_{\rm{Dirac}} = - \int d^4\tilde{x}\ i\bar{\tilde{\psi}} \left(\Gamma^{{r}} \partial_{\tilde{r}} + \Gamma^{{t}} (\partial_t - iA_t) +  \frac{\Omega}{\Sigma} \Big(\Gamma^{{x}}\partial_x + \Gamma^{{y}} (\partial_y - iBx) \Big) - m\Omega \right) \tilde\psi\nn
\ee
Note that the kinetic terms in the $\tilde{r}$ and $t$ directions are those of a fermion in a flat space. 

\para
 Because $\Omega$ and $\Sigma$ depend on radial position only, the resulting Dirac equation for $\tilde \psi$ is separable: one can solve for the $x$ and $y$ parts of the separable solution and write the general solution as a superposition of all such modes. This provides the familiar Landau level decomposition of the bulk fermions, 
 \be
\tilde\psi = \int \frac{dk}{2\pi} \sum_{n=0}^\infty e^{-iky} \left( X_{n-1}(x,k) P_+ + X_{n} (x,k) P_- \right) \tilde\psi_{n,k} (\tilde{r},t).
\label{decomp}\ee
The  matrices $P_\pm$ are projections onto spin-up and spin-down degrees of freedom.
\be
P_\pm = \frac 1 2 (1 \pm i \Gamma^{{x}} \Gamma^{{y}} )
\ee
while the profile functions $X_n(x,k)$ are the usual, orthonormal wavefunctions of the harmonic oscillator expressed in terms of Hermite polynomials $H_n$,
\be
 X_n (x,k) & = & \left( \frac{\sqrt{B}} {2^n n! \sqrt{\pi}}\right)^{\frac 1 2} \exp\left(-\frac {B} 2 \left( x + \frac k{B}\right)\right) H_n \left( \sqrt{B} \left( x + \frac k {B} \right)\right)\ \ \ \ n\geq 0 \nn\ee
and  $X_{-1}\equiv 0$. 
%
%
Each $n=0,1,2,\ldots$ labels a separate relativistic Landau level. Meanwhile, the degeneracy within each level is captured by the integral over $k$. Individual  states  labelled by $k$ are not  translationally invariant in both $x$ and $y$ directions. However, translationally invariant states can be constructed by taking a superposition of all $k$ states. Because, by definition, $X_{-1} \equiv 0$, the $n=0$ Landau level contains only spin-down states.  This reflects the fact that we mentioned earlier: the lowest Landau level is spin polarised and contains only half the states of higher levels. 

\para
It is common practice to replace the integral over degenerate $k$-modes with a discrete sum. This is achieved this by temporarily restricting the system to a finite transverse area $A$. It is  simple to show that the degeneracy of each spin state in a Landau level is $BA/2\pi$, and we thus make the replacement
\be
\int \frac{dk}{2\pi}\ \rightarrow\ \sum_k^{BA/2\pi}.
\ee
Within each Landau level, the remaining dynamics then takes place only in the radial, $\tilde{r}$, direction.  We emphasise this by introducing two-component spinors, $\xi_+$ and $\xi_-$, representing individual spin-up and spin-down degrees of freedom respectively,
\be
\xi_{-\,n,k}(\tilde{r},t) = P_- \,\tilde{\psi}_{n,k}(\tilde{r},t), \qquad \xi_{+\,n,k}(\tilde{r},t) = i \Gamma^5 \Gamma^{y} P_+ \,\tilde{\psi}_{n,k}(\tilde{r},t).
\ee 
To accompany these, we define two-component gamma matrices 
\be
\gamma^r = \Gamma^{r} P_-, \qquad \gamma^t = \Gamma^{t} P_-, \qquad \gamma^3 = \Gamma^5 P_-
\ee
which satisfy the $1+1$-dimensional Clifford algebra $\{ \gamma^\mu, \gamma^\nu \} = 2\eta^{\mu\nu}$, with  $\gamma^3 = \gamma^t \gamma^r$. 
%
%
%
%

\para
After this whirlwind of redefinitions, we can finally write our $d=3+1$ dimensional bulk fermion action as a sum over effective $d=1+1$ dimensional actions for spin-up and spin-down fermions in each Landau level. Substituting the solution \eqn{decomp} into the bulk action yields
\be
S_{\rm{Dirac}} = - \sum_k^{BA/2\pi} \int d\tilde{r} dt && \left[ \sum_{n=0}^\infty \ i\bar\xi_{-\,n,k} (\gamma^r\partial_{\tilde{r}} + \gamma^t (\partial_t - iA_t) - m\Omega)\xi_{-\,n,k} \right. \nn \\
&& + \sum_{n=1}^\infty \  i\bar\xi_{+\,n,k}  (\gamma^r\partial_{\tilde{r}} + \gamma^t (\partial_t - iA_t) - m\Omega)\xi_{+\,n,k} \nn\\
&& + \left. \sum_{n=1}^\infty \sqrt{\frac{2B\Omega^2n}{\Sigma^2}} \left( i\bar\xi_{+\,n,k} \gamma^3 \xi_{-\,n,k} - i\bar\xi_{-\,n,k} \gamma^3 \xi_{+\,n,k}\right)\right]
\nn\ee
We stress again that the lowest Landau level, $n=0$, contains only spin-down $\xi_-$ states. Notice that, for the lowest Landau level, the only effect of the curved spacetime lies in the position dependent mass in the first term, capturing nothing more than the familiar gravitational red-shift. The last term in the above expression describes the energy cost to excite a mode in a higher Landau level. 


\subsection{Bosonization of the Lowest Landau Level}


Our focus in this paper is on dynamics of the bulk fermions at energies $E< \sqrt{B}$. In this regime, all higher $n\geq 1$ Landau levels decouple and an effective dimensional reduction takes place, with the fermion dynamics well captured by a simple $d=1+1$ dimensional action for the lowest Landau level.  The effective dimensional reduction of the lowest Landau level plays an important role in number of settings including, for example, the phenomenon of magnetic catalysis. (See \cite{mcreview} for a recent review).

\para
Since we focus on the lowest Landau level, from now on we will  drop both the $n=0$ label and spin $-$ label from the associated spinor and write simply $\xi_k\equiv \xi_{-\,0,k}$. Moreover, because we are interested in translationally invariant solutions, we are at liberty to also write the Maxwell action in a similarly dimensionally reduced form. The coupled system of fermion and gauge fields is then described by the action
\be S= S_{\rm LLL} + S_{\rm Maxwell}\label{lllm}\ee
with the lowest Landau level fermions governed by 
\be S_{\rm LLL} = -  \sum_k^{BA/2\pi} \int d^2x\ i \bar\xi_k (\gamma^\mu \partial_\mu - i\gamma^\mu A_\mu - m\Omega)\xi_k   \label{lll}\ee
and the gauge fields dictated by a two-dimensional Maxwell action in curved space which, for reasons that will become apparent later, we normalize as 
\be
S_{\rm Maxwell} = - \frac{BA}{2\pi} \int d^2x \left( \frac{1}{4g^2} F_{\mu\nu} F^{\mu\nu} + \frac{\pi\Omega^2}{e^2\Sigma^2} B\right)\label{maxwell}
\ee
In both actions \eqn{lll} and \eqn{maxwell},  $\mu$ and $\nu$ indices run over $\tilde{r}$ and $t$ only and indices are raised with the 2D Minkowski metric $\eta^{\mu\nu}$. Similarly the integral $d^2x\equiv d\tilde{r}dt$.  Notice that, viewed as an action governing the 2d gauge field $A_\mu$, only the first term is relevant. The second term, involving the magnetic field, plays a role only when determining the gravitational backreaction. We have introduced the effective two-dimensional coupling constant
\be \frac{1}{g^2(r)}=\frac{1}{e^2B}\frac{2\pi\Sigma^2}{\Omega^2}\label{g2}\ee
From \eqn{thoseguys}, we have $\Sigma^2/\Omega^2=1/f(r)$, the black hole emblackening factor. This means that $g^2$ is constant in an AdS spacetime, but approaches zero close to a black hole horizon.

\para
The action \eqn{lllm} captures only the interaction between fermions and gauge fields; we will discuss the backreaction of these fields on the metric itself shortly. However, \eqn{lllm} is very familiar: it is  the Schwinger model in $d=1+1$ curved spacetime, albeit with a large degeneracy, $BA/2\pi$, of identical species of massive fermions. As observed previously in \cite{eboli}, when written in flat space, with conformally rescaled fermions, the effect of the spacetime curvature is captured by a position-dependent effective mass $m\Omega$ and a position-dependent effective gauge coupling $g^2$.

\subsubsection*{Bosonization in Curved Spacetime}

Fermions in two dimensions have a magical property: they can be treated as bosons. In flat space, a two-component Dirac fermion operator $\xi$ maps into a real scalar operator $\phi$ (see, for example, \cite{senechal,shankar} for pedagogical introductions).  The kinetic terms of each field are related by 
\be i\bar{\xi}\gamma^\mu\partial_\mu\xi = \frac{1}{8\pi}\partial_\mu\phi\partial^\mu\phi\label{kin}\ee
The vector current of the fermion maps onto a topological current of the scalar
\be
\bar{\xi}\gamma^\mu\xi  = \frac{1}{2\pi} \epsilon^{\mu\nu}\partial_\nu\phi\label{vector}\ee
Meanwhile, the axial symmetry of the fermion becomes a shift symmetry of the scalar, $\phi\rightarrow \phi + c$, with the currents related by
\be \bar{\xi}\gamma^3\gamma^\mu\xi = \frac{1}{2\pi}\partial^\mu\phi\label{axial}\ee
As we will see shortly, this has an interesting consequence for the 2d and, for us, the 4d anomaly. 
Finally, the fermion mass term becomes a sine-Gordon potential in the scalar theory,
\be im\bar{\xi}\xi = \frac{m\Lambda}{\pi}\cos\phi\label{gordonsalive}\ee
which clearly breaks the shift symmetry, as it should. Here $\Lambda$ is a regularization scale, into which many subtleties of the bosonization map are swept. We shall discuss some of these below. This regularization scale is needed in the definition of the massless $\phi$ field since the propagator in two dimensions is a logarithm. It also appears in the definition of normal ordered products of fields which we have implicitly used in the expressions above. As we shall describe in more detail later, when computing physical quantities using the semi-classical bosonized action, $\Lambda$ should be identified with an appropriate physical scale in the problem. 

\para
For our purposes, there are two  further issues that arise in the bosonization procedure. The first is because we are working in a curved spacetime. This means that all energy scales -- and this includes the regularization scale $\Lambda$ -- must be appropriately red-shifted as we move in space. For this reason, we should replace $\Lambda$ in \eqn{gordonsalive} with\footnote{This observation seems to have been missed in a number of earlier papers on bosonization in curved spacetime.  A correct discussion can be found in \cite{urban}. Failure to make the regulator position dependent in this manner results in a number of pathologies including, as previously pointed out in \cite{dorn}, a breakdown of diffeomorphism invariance and, relatedly, the inability to construct a 4d stress tensor. Moreover, bosonization in AdS spacetime only respects the scaling symmetry if the red-shift factor is correctly treated. (For us, this scaling symmetry is manifest in the equation of motion \eqn{did1} for the scalar).} $\Lambda \rightarrow \Lambda \Omega(\tilde{r})$. 

\para
The second issue involves the fact that we have a large number, $BA/2\pi$, of fermions that we wish to bosonize. These are labelled by the index $k$ in \eqn{lll}. This results in an equally  large number of scalar fields $\phi_k$. However, individually, each of these states breaks translational invariance in the $x-y$ plane. Translationally invariant solutions only arise if each of these bosons  moves as one. For this reason, we identify each $\phi_k\equiv \phi$. The net result is that,
 when restricted to  solutions which are translationally invariant in the $x-y$ plane, the action \eqn{lll} is equivalent to 
\be
S_{LLL} = - \frac{BA}{2\pi}  \int d^2x\ \left( \frac 1 {8\pi} \partial^\mu \phi \partial_\mu \phi + \frac{m\Lambda \Omega^2}{4\pi} (1-\cos \phi) + \frac 1 {4\pi}  \epsilon^{\mu\nu} \phi F_{\mu\nu} \right) \label{bosoned}\ee
Here, $\epsilon^{tr} = -\epsilon^{rt} = +1$. As promised, we will say more about the relevant scale $\Lambda$ in due course. 

\para
The idea that, in certain circumstances -- often involving the presence of a magnetic field -- it is profitable to bosonize fermions in higher dimensions is not new. Perhaps the first application  -- and the one which provided inspiration for the current work -- is the solution of s-wave scattering of fermions off a magnetic monopole due to Callan \cite{callan2}.  Indeed, the s-wave fermi zero mode is morally equivalent to the lowest Landau level states that we consider here\footnote{This analogy can be seen by starting with a magnetically charged black hole in global AdS. Upon taking the limit to the planar Reissner-Nordstrom in the Poincar\'e patch, the s-wave zero modes evolve to the lowest Landau level.}. 
Yet more recently, there has been a suggestion to use 2d bosonization in conjunction with Eguchi-Kawai reduction to study the planar limit of certain 3d field theories \cite{aleksey}. There was also a recent interesting proposal for a genuine bosonization of fermions coupled to Chern-Simons terms in $d=2+1$ dimensions \cite{ofer}.

\subsection{Fermions, Kinks and  Anomalies}
 
In the following section we will study solutions to \eqn{bosoned} in asymptotically AdS spacetimes in some detail and explore their implications for the boundary field theory. Here we just make a few general comments on the physics captured by this action, all of which is standard fare in the bosonization literature. This will also shed some light on the meaning of the scale $\Lambda$. 

\para
First, let us ignore the coupling to the gauge field and focus only on the first two terms in \eqn{bosoned}. The potential admits vacua at each $\phi=2\pi n$, for $n\in{\bf Z}$. Kinks which interpolate between adjacent vacua have unit topological charge and hence, from \eqn{vector}, are identified with the fermions of the original theory. This identification also tells us that --- in the absence of any gauge interaction --- we should identify $\Lambda \sim m$, so that the mass of the fermion matches that of the kink. 
When viewed from the perspective of our original theory in AdS$_4$, the kinks are domain walls of fermionic charge, translationally invariant in the $x-y$ direction. They have energy  $\sim BAm\Omega/2\pi $. The interpretation of this is clear: the domain walls describe $BA/2\pi$ bulk fermions, each of mass $m$, suitably redshifted. We will later identify these as the filled lowest Landau levels of various carrier bands in the dual theory.

\para
Now let's see how this story is affected by the coupling to the gauge field. The equation of motion for the gauge field shows us that the kinks are electrically charged as expected,
\be
\partial^\mu \left( \frac{1}{ g^2(r)} F_{\mu\nu}\right)  =- \frac {1} {2\pi} \epsilon_{\mu\nu} \partial^\mu \phi\ee
This is where issues of the regularization scale $\Lambda$ become subtle since, in some contexts, it is necessary to take $\Lambda$ proportional to the gauge coupling $g$ when treating \eqn{bosoned} semi-classically.  In the  next two paragraphs we will explain why we shouldn't do this and must instead  retain $\Lambda\sim m$. Readers uninterested in these subtleties are invited to jump to equation \eqn{goodagain}.

\para
Perhaps the simplest, and most compelling, way to determine the regularization scale $\Lambda$ is to use the fact that we're working in a higher dimensional, curved spacetime. 
 In Appendix \ref{appa}, we construct the fermionic contribution to the $d=3+1$ stress tensor in terms of the bosonized field $\phi$. This stress tensor depends explicitly on $\Lambda$ and  the requirement that is covariantly conserved puts strong constraints on what $\Lambda$ can be. This is essentially because the effective gauge coupling $g^2$, defined in \eqn{g2}, depends on the radial position on a way which remembers its four-dimensional origin.  We show that the only consistent choice of regularization scale is $\Lambda\sim m$.

\para
It is also useful to address the issue directly in a two-dimensional language. In the usual discussion of the Schwinger model in flat space, one puts  $\Lambda \sim g$ rather than $\Lambda \sim m$. (At least this is true in the regime $g^2\gg m^2$ where the bosonized action is weakly coupled). Here the choice of $\Lambda$  reflects the fact that the mass of the electron is renormalized by the interaction with the gauge field, as can be seen in the bosonic picture by integrating out the gauge field. However, even here one should only put $\Lambda \sim g$ for the Schwinger model with a single species of fermion. In the presence of two fermion species, $\xi_1$ and $\xi_2$, the story changes. While the singlet boson $\phi_1 + \phi_2$ acquires an effective mass proportional to $g$, the other boson combination $\phi_1 - \phi_2$ acquires a mass proportional to $m^{2/3} g^{1/3}$. A careful treatment shows that the correct normal ordering scale is now $\Lambda \sim m^{1/3}g^{2/3}$ \cite{coleman2}.
Extending this argument to $N_f$ species of fermions, the normal ordering scale becomes $\Lambda^{N_f+1} \sim  m^{N_f-1}g^{2}$ 
\cite{smilga}. Since we have $BA/2\pi$ species of fermions, we should take the $N_f\rightarrow \infty$ limit. We again find $\Lambda \sim m$ as promised\footnote{There is yet a third way to fix the regularization scale $\Lambda$ which, moreover, provides insight on a slightly different topic. One physical quantity in which the regularization scale appears is the chiral condensate $i\langle \bar{\xi}\xi\rangle = \Lambda/\pi$ and, in  the lowest Landau level, it is known that the chiral condensate is proportional to the mass (see, for example, \cite{dunne}). Furthermore, with the identification $\Lambda \sim m$, it is simple to show that a number of higher dimensional phenomena which rely on the lowest Landau level -- such as the chiral magnetic effect and the chiral spiral (see again \cite{dunne} for a review) -- follow trivially from the equations of motion of the bosonized theory.}.

\para 
The arguments above for  $\Lambda \sim m$ do not fix an overall, order one coefficient. For this reason, we'll replace $m\Lambda \rightarrow m_R^2$, the renormalized mass, in the above expressions. The equation of motion for $\phi$ then reads 
\be
\partial^\mu \partial_\mu \phi &=& m_R^2 \Omega^2 \sin \phi + \epsilon^{\mu\nu} F_{\mu\nu}
\label{goodagain}\ee
The second term on the right-hand side is particularly interesting. It captures the axial anomaly in two dimensions, as can be seen by substituting the expression \eqn{axial} for the axial current. However, our theory has a four dimensional origin and all expressions can be rewritten in in terms of the four dimensional fermion $\tilde{\psi}$. Setting $m=0$ in \eqn{goodagain} and taking into account the degeneracy, $BA/2\pi$, of the lowest Landau level, we see that the two-dimensional anomaly is, at heart, the four dimensional  anomaly\footnote{Here we include only the contribution from the lowest Landau level. Higher Landau levels transform as $\xi_{n\pm} \to e^{\mp i \theta \gamma^3 / 2}\xi_{\pm}$ under the chiral rotation $\psi \to e^{i\theta \Gamma^5 / 2} \psi$. The axial anomalies from spin-up and spin-down states of the higher Landau levels cancel out, leaving only the contribution from the lowest Landau level.}
\be
\int dx dy\ \dot{Q}_{\rm axial} = \int dxdy\ \frac{ \vec{E}.\vec{B}}{2\pi^2}
\nn\ee
where the axial charge is $Q_{\rm axial}=\tilde{\bar\psi} \Gamma^{t} \Gamma^5 \tilde{\psi}$. 
The axial anomaly is a property of the quantum theory of the fermion field. It is a beautiful feature of bosonization that the classical equations of motion capture this quantum aspect of fermions.

\subsubsection*{Gravitational Backreaction}

So far, we have restricted our discussion to the dynamics of fermions and gauge fields in the background of a fixed, albeit arbitrary, asymptotically AdS$_4$ background. Ideally, one would like to generalise this to include backreaction on the metric itself and, indeed, in the framework of bosonization there is neither  conceptual nor technical obstacle to this. In appendix \ref{appa}, we compute the four-dimensional stress tensor arising from fermions in the lowest Landau level and reduce the Einstein-Maxwell-Dirac equations to a set of coupled, ordinary differential equations. 

\para
Nonetheless, in this paper we do not solve the full backreacted equations. The reason for this is rather trivial: in the regime, $B \gtrsim \mu^2$ in which we are working, the backreaction on the metric is dominated  by the magnetic field $B$. Furthermore,  while the electron star screens the electric field in the bulk, it does nothing to change the magnetic field. The upshot of this is that, at least in the regime where the most interesting things happen, the fully back-reacted solutions do not differ substantially from the purely magnetic \RN \ black hole background, described by \eqn{rn1} and \eqn{rn2} with $\mu=0$. For this reason, in the following section we use bosonization to study the lowest Landau level of electron stars in the purely magnetic \RN \ black hole.

\section{Phenomenology of the Lowest Landau Levels}\label{3sec}

In this section we describe in some detail the solutions to our bosonized system and their implications for the dual boundary field theory. We will focus entirely on equilibrium properties but will, nonetheless, find a rich phenomenology. 

\para
One of the main obsessions in this section is on the charge density, $\rho$, of the boundary field theory. As always, this has the interpretation of the bulk electric field near the boundary. More precisely, working in $A_r=0$ gauge, the charge density can be read off from boundary behaviour of $A_t$,
\be A_t\rightarrow -\mu + e^2\rho r\qquad {\rm as}\  r\rightarrow 0\label{24}\ee
From this we see that the charge density $\rho$ is proportional to the value of the electric field $F_{tr}$ on the boundary. This charge density can have two qualitatively different sources \cite{fraction1,fraction2}. Firstly, the electric field can be sourced by charged matter in the bulk. For us, this charged matter means the fermions or, after bosonization, the kinks of the scalar field. In the boundary theory, this corresponds to charge carried by gauge invariant fermionic operators $\Psi$ dual to the bulk field $\psi$. Such charge is referred to as `mesonic' or `cohesive'. As shown in \cite{sachdev}, this charge contributes to the Luttinger count. 

\para
An alternative source of electric field is provided by the horizon of a \RN\ black hole. In the boundary theory, such charge is carried by the elementary, non gauge-invariant fields of the large N gauge theory (whatever they may be). Such  charge is said to be `fractionalised'\footnote{A particle theorist would call such charge deconfined since particle theorists are used to having fundamental non-Abelian gauge fields at a high scale. The term fractionalised is more appropriate in the condensed matter setting where non-Abelian gauge fields first emerge at some low scale then de-confine in  a non-trivial manner, the net effect being that  elementary particles -- such as the electron -- appear to fractionalise.}. 

\para
Much of the focus in the story of electron stars is to determine which of the two sources of charge -- cohesive or fractionalised -- is the stable ground state in any given setting. Indeed, one can find situations where both play a role  \cite{hartnoll2}.  The main purpose of this section is to describe how these two types of charges arise in the presence of a strong magnetic field. We will find that there is an interesting interplay between the two. 

\para
Although we are ultimately interested in the physics in the background of a bulk \RN\ black hole, we will first start in a simpler setting where we can better illustrate the main ideas of formation of an electron star in the framework of bosonization. In Section \ref{bhsec}, we will place our electron stars in a black hole spacetime and illustrate a number of different phenomena that occur in the presence of a horizon.  

\subsection{Electron Stars in a Hard wall}\label{hwsec}

To describe the formation of an electron star, we start with the simple situation of a AdS hard wall. The metric is
\be
ds^2 = \frac{L^2}{r^2}( -dt^2 + dr^2 + dx^2 + dy^2 )
\ee
The UV boundary lies at $r=0$. In the infra-red, we assume that the space stops abruptly at $r=r_\star$. This is the the hard wall.  This geometry was also used in \cite{sachdev} in the construction of a quantum electron star. 

\para
We thread this geometry with a constant magnetic field $F_{xy}=B$.  For the purposes of this toy model, we will assume that this magnetic field is emitted by the hard-wall. Furthermore, we ignore the back-reaction of the magnetic field on the metric. (Such ignorance is justified if $Br^2_\star \ll  \sqrt{e^2L^2/\kappa^2}$).

\para
Already in vacuum, there are  a number of interesting effects experienced by fermions in a the magnetic hard wall geometry \cite{magcat,magcat2}. Here we differ from these papers by turning on a charge density in the boundary field theory. This means that, close to the boundary, the bulk gauge field takes the form \eqn{24} with $\rho\neq 0$. We will further insist that no electric field is emitted from the hard-wall; the electric field must therefore be entirely sourced  by bulk fermions which, after bosonization, means kinks of the scalar field $\phi$. We would like to ask what charge densities $\rho$ can be built in this manner.

\para
In fact, this is a variant of a classic problem in physics. If you stand on your head, the kinks which source the electric field can equally well be thought of as screening an electric field emitted from the boundary. This is the question of charge screening in the Schwinger model. Famously, after bosonization this classic problem becomes a classical problem\footnote{This statement is almost true. The bosonized theory can be treated semi-classically in the regime $g^2\gg m_R^2\Omega^2$ where its self-interactions are weak. This translates to the requirement $Be^2 \gg m^2L^2/r^2$. Notice that, for $m\neq 0$,  a semi-classical treatment of the bosonized action is not strictly valid close to the boundary. As we shall see, at least for small masses the most interesting physics occurs suitably far from the boundary.} \cite{coleman}. 

\para
Working in Coulomb gauge, and assuming a time independent configuration, the classical equations of motion of the scalar $\phi$ and gauge field are
\be
\partial_{r}^2 A_t  =  - \frac{ e^2 B}{4\pi^2} \partial_{r} \phi \ \ \ {\rm and}\ \ \ \  
\partial^2_{r} \phi  =  \frac{m_R^2L^2}{r^2} \sin \phi - 2\partial_{r}A_t
\label{do1}\ee
The first of these equations, Gauss' law,  can be integrated trivially to give an expression for the electric field in terms of the scalar field.
\be
\partial_{r}A_t = -\frac{ e^2 B}{4\pi^2} (\phi - \phi_0)\label{done1}
\ee
Here we have introduced an integration constant, $\phi_0$. To understand the meaning of this integration constant, we need to discuss boundary conditions for $\phi$ at $r=0$. These are inherited from the boundary conditions  \eqn{fermibc} for the fermions\footnote{In fact, there appear to be interesting choices in the allowed boundary conditions for both fermions and scalars. We have relegated the details to Appendix \ref{appbound}.} and are simply $\phi=0$ \cite{callan2}.  Comparing \eqn{done1} to \eqn{24}, we learn than $\phi_0$ determines the charge density
\be
\rho = \frac{B}{4\pi^2} \phi_0
\label{likethis}\ee
In the context of free electrons, $\phi_0/2\pi$ is called the {\it filling fraction}; it tells you how many Landau levels are filled. In our case, we are already restricting ourselves to  the lowest Landau level but, of course, we are describing a strongly interacting, large $N$ theory. We will continue to refer to $\phi_0/2\pi$ as the filling fraction. However, it will now have a slightly different interpretation, telling us how many bands -- or different species -- of fermions have their lowest Landau level filled\footnote{A comment on the regime of validity: we cannot increase $\phi_0$ too much or we will start to fill higher Landau levels and our framework of bosonization will break down. We are parameterically safe, with $\mu<\sqrt{B}$, provided that  $\phi_0 < 1/e$. This means that we can fit up to $n\sim 1/e$ kinks in our geometry.}. 

\para
Substituting \eqn{done1} into the second equation in \eqn{do1}, we arrive at a second order differential equation for the scalar,
\be 
\partial^2_{r}\phi = \frac{m_R^2L^2}{r^2} \sin \phi + \frac1{r_s^2} (\phi - \phi_0)
\label{did1}\ee
where we have introduced a new length scale, 
\be
r_s^2 = \frac{2\pi^2}{ e^2 B},
\label{rs}\ee
This  represents the characteristic length scale over which the bulk electric field is screened (at least in the case of small mass). 

%

\para
Before solving \eqn{did1}, we must first discuss the boundary conditions which we impose at the hard wall. The canonical choice is to again impose $(1-\Gamma^r)\psi=0$ at $r=r_\star$, just  as we did at the boundary \eqn{fermibc}. In the UV, this translated into the requirement $\phi=0$ and one might think that that we should also insist on  this at the hard wall. 
But this is not quite right. The boundary condition \eqn{fermibc} is invariant under a chiral rotation of $2\pi$. Yet, under such a rotation, $\phi \rightarrow \phi +2\pi$ This means that the boundary condition on the fermion actually allows a discrete choice of boundary conditions on the scalar, namely 
\be \phi = 2\pi n, \ \ n\in {\bf Z}\ \ \ \ \ \ {\rm at} \ r=r_\star\label{29}\ee
The one further requirement that we need to impose is that the hard wall emits no electric field. In the bosonized language, the relationship \eqn{done1} tells us that this becomes $\phi=\phi_0$ at $r=r_\star$. Combined with \eqn{29}, we learn that  the hard wall background admits translationally invariant electron stars in the lowest Landau level  only for integer valued charge density
\be \rho = \frac{B}{2\pi}n\nn\ee
The same result was found in \cite{koji} which examined the Luttinger count for electron stars in a hard wall and magnetic field. 
If one tries to impose a non-integer filling fraction, the resulting state is presumably not translationally invariant in this setting. We will see shortly that this situation changes in the presence of a black hole horizon. (In Appendix \ref{appbound} we also describe boundary conditions at the hard wall which result in fractional charge density).

\para
For massless bulk fields, $m=0$, the solution to \eqn{did1} is a simple exponential 
\be \phi = \phi_0\,\frac{1-e^{-r/r_s}}{1-e^{-r_\star/r_s}}\label{exprofile}\ee
This describes the profile of fermion charge density in the bulk sourcing the electric field. 

\begin{figure}
\resizebox{80mm}{!}{\includegraphics{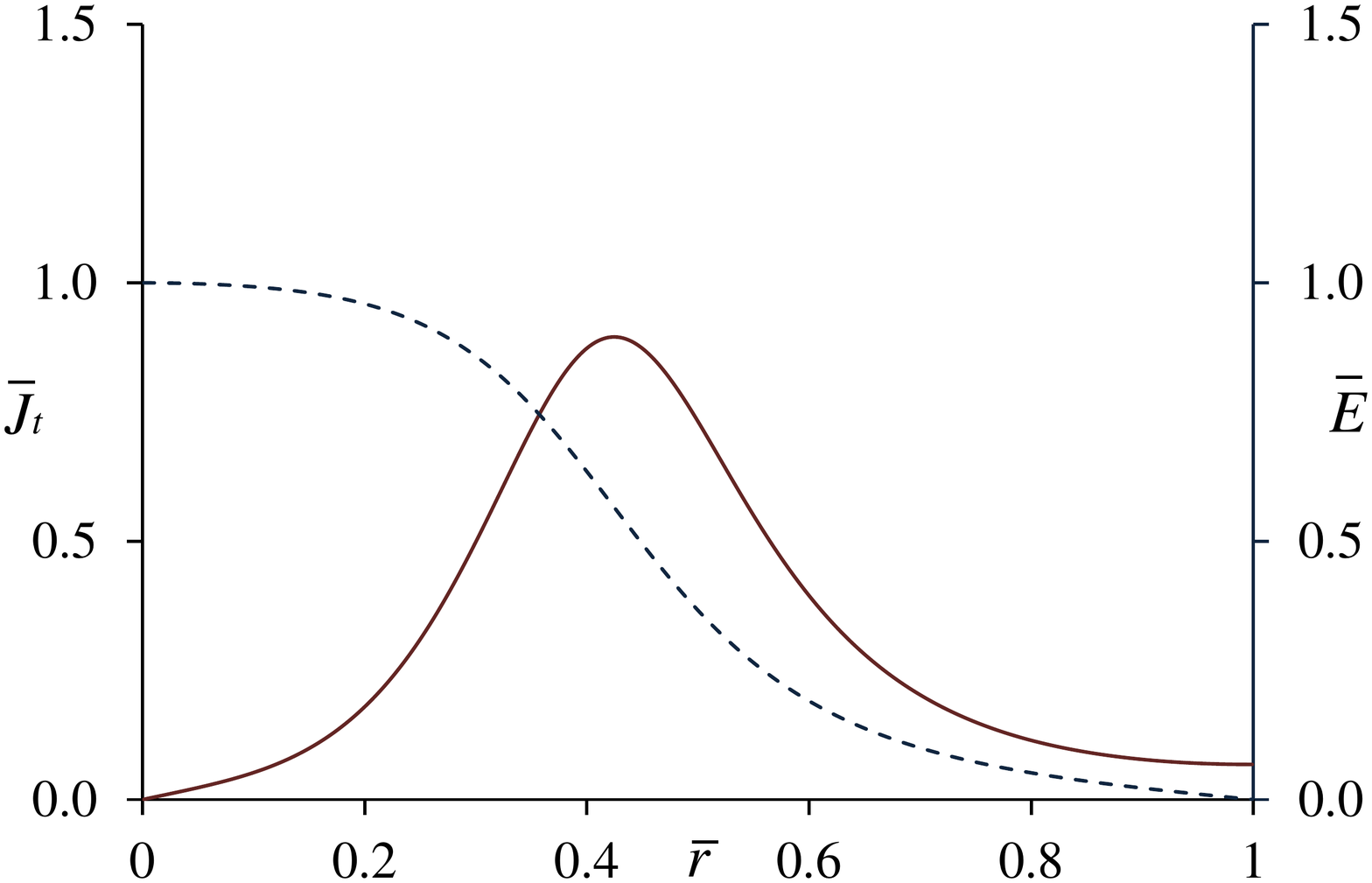}}
\resizebox{80mm}{!}{\includegraphics{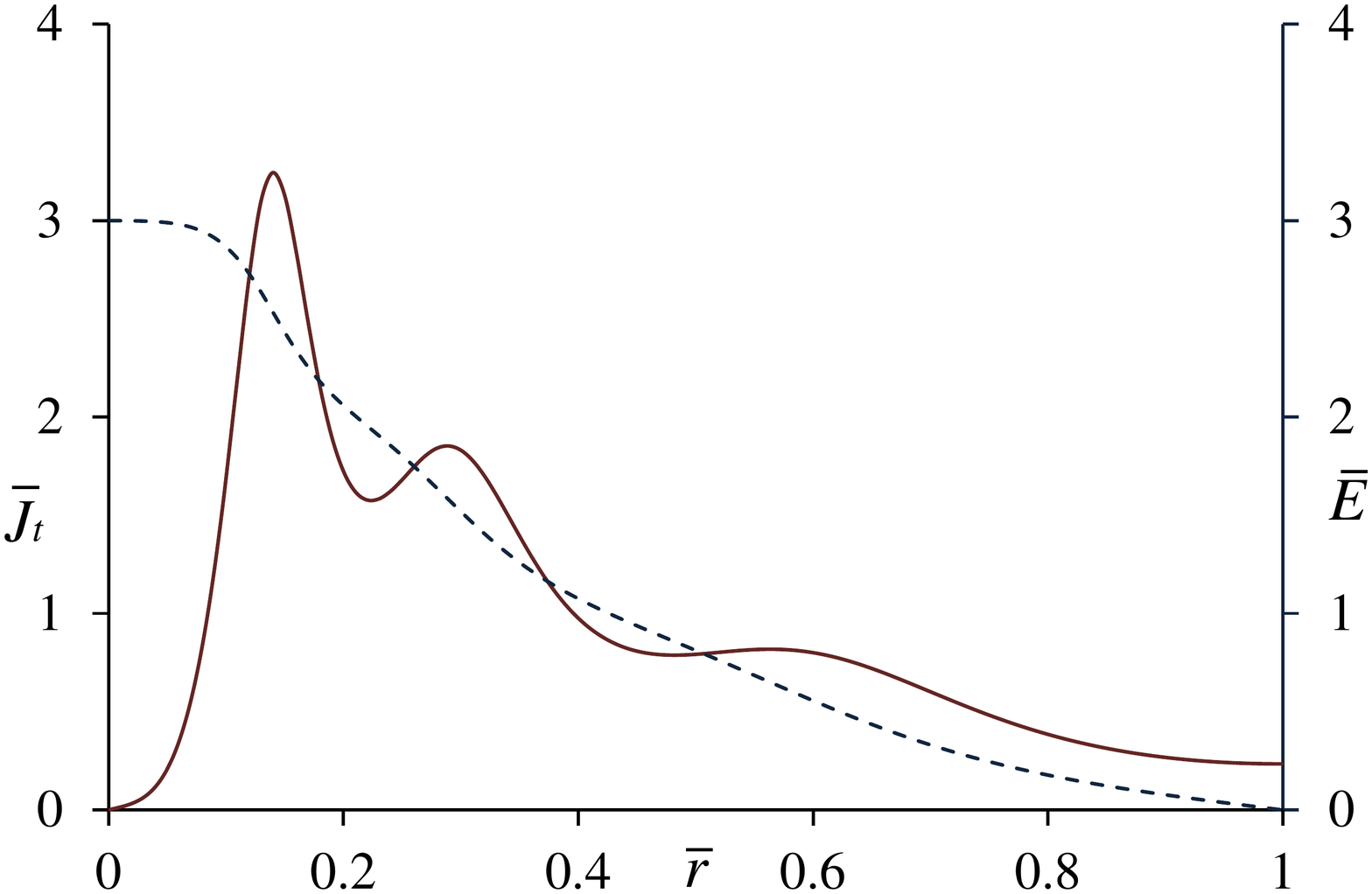}}
\caption{Electron stars in a hard wall. The solid line shows the bulk fermionic charge density, normalised as $\bar{J}_t=r_s\partial_r\phi/2\pi$ as a function of radial position $\bar{r}= r/r_\star$; the dotted line shows the bulk electric field, again normalised as $\bar{E}= r_s^2F_{rt}/\pi$. The left hand plot has filling fraction $\phi_0/2\pi =1$ with just a  single kink; the right-hand plot has  $\phi_0/2\pi=3$ and three distinct peaks. Both of these plots are made with $m^2_RL^2 = 16$ and $r_s^2/r_\star^2 = 1/10$.}
\label{hardfig}
\end{figure}

\para
For $m\neq 0$, we solve \eqn{did1} numerically\footnote{We use a fourth-order Runge-Kutta algorithm with variable step size. The IR boundary condition is implemented using a shooting method.}. The resulting profiles are shown in Figure \ref{hardfig}.  Since the mass makes it unfavourable to place fermions near the UV, the charge density of the electron star is pushed further away from the boundary as the mass increases. Most strikingly, the electron star develops kinks of localised charge density within its profile. These are simply the kinks of the sine-Gordon model, now transplanted into AdS.  If you allow more fermions in each Landau level by increasing $B$, then the boundary electric field increases and the kinks are drawn towards the UV boundary.

\para
Within each kink, $\phi$ increases by  $2\pi$ and, using \eqn{do1}, therefore sources a charge density in the boundary of $B/2\pi$. This is the charge density obtained by filling a  lowest Landau level of states. From the perspective of the boundary field theory, we are filling the lowest Landau levels  of different species -- perhaps a better name is different carrier bands -- of fermions. In the bulk, these bands are associated to different radial harmonics of the fermion field\footnote{The presence of different carrier bands is also seen in the original electron star papers where it results in concentric rings of Fermi surfaces \cite{sean3,hong}. Indeed, in the limit $mL\gg1 $ necessary to treat the electron star as a fluid, these Fermi surfaces are densely packed. The same phenomenon occurs here: the width of a kink at position $r$ is roughly $r/m$. As $m$ increases, the kinks become sharper and more closely spaced.}. Note, however, that unlike in the free theory, the Landau levels are not delta-functions in energy space. They have been broadened by the strong interactions of the boundary theory, captured by the AdS geometry.

\para
Viewing the radial coordinate $r$ in AdS as an energy scale, $E\sim 1/r$, it is tempting to view the  kink profiles in Figure \ref{hardfig} (and those of Figure \ref{bhfig}) directly as a plot of the density of states of the lowest Landau level of the boundary field theory.

\subsection{Black Holes}\label{bhsec}

We now turn to the main case of interest: electron stars in the background of a magnetic \RN\ black hole. We take the geometry to be 
\be ds^2 = \frac{L^2}{r^2}\left(-f(r)dt^2+\frac{dr^2}{f(r)}+dx^2+dy^2\right)\nn\ee
with 
\be f(r) = 1-\frac{4r^3}{r_h^3}+\frac{3r^4}{r_h^4}\nn\ee
The position of the horizon is fixed by the magnetic field, 
\be r^2_h=\sqrt{\frac{6e^2L^2}{\kappa^2}} \frac{1}{B}\nn\ee
Our use of this background requires some explanation. As we shall see, we will be interesting in solutions containing both cohesive charge, carried by the star, and fractionalised charge lying behind the horizon. Yet we do not allow the position of the horizon to depend on the electric charge of the black hole. This sounds inconsistent. In fact, as we show in Appendix \ref{appa}, the backreaction of the electric field on the position of the horizon is negligible provided $e^2\phi_0\ll1 $ and $e^2\phi_0m_RL\ll 1$. It is easy to understand why. Even in the UV,  the magnetic field $B$ is larger than the electric field. By the time we get down to the horizon the vast majority of the electric field has been screened by the star and what remains does not meaningfully affect the position of the horizon.

\para
The equations of motion again tie the bulk electric field to the scalar,
\be 
\partial_{r}A_t = -\frac{ e^2 B}{4\pi^2} (\phi - \phi_0)
\label{doing1}\ee
Meanwhile, the second order equation governing the scalar itself is now given by
\be
\partial_{r}(f(r)\partial_{r}\phi) = \frac{m_R^2L^2}{r^2} \sin \phi + \frac1{r_s^2} (\phi - \phi_0)
\label{does1}\ee
where, as in the previous section, the screening length is defined as  $r_s^2 = {2\pi^2}/{ e^2 B}$.

\para
The primary difference with the hard wall background lies in the boundary conditions imposed on $\phi$. In the UV, we again set $\phi=0$ at $r=0$. As before, this allows us to relate $\phi_0$ to the charge density: $\rho = { B\phi_0}/{4\pi^2} $. However, we do not impose Dirichlet boundary conditions in the IR. Instead we require only that $\phi$ and $\partial_r\phi$ are regular on the horizon. 
However, because $f(r)$ has a double zero in the extremal \RN\ black hole, this in turn requires that $\phi(r_h)$ lies in an extremum of the effective potential\footnote{A full analysis of the requirements in the near horizon AdS$_2\times {\bf R}^2$ regime is given in Appendix \ref{appb}.}
\be
V_{IR}(\phi) = m_R^2L^2(1-\cos\phi) + \frac{ r_h^2}{2r_s^2} (\phi-\phi_0)^2.
\label{vir}\ee

\begin{figure}[!h]
  \resizebox{80mm}{!}{\includegraphics{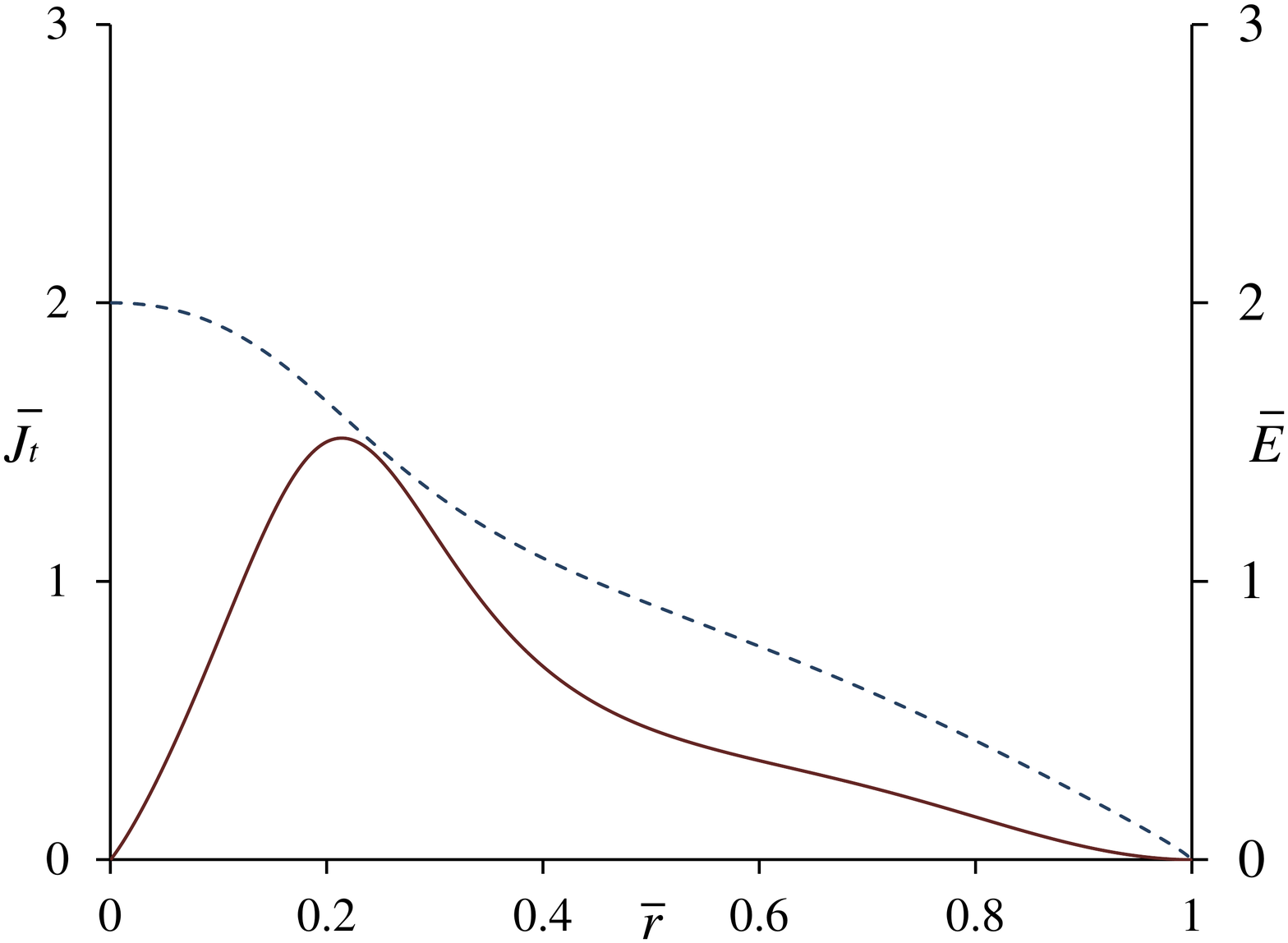}}
\resizebox{80mm}{!}{\includegraphics{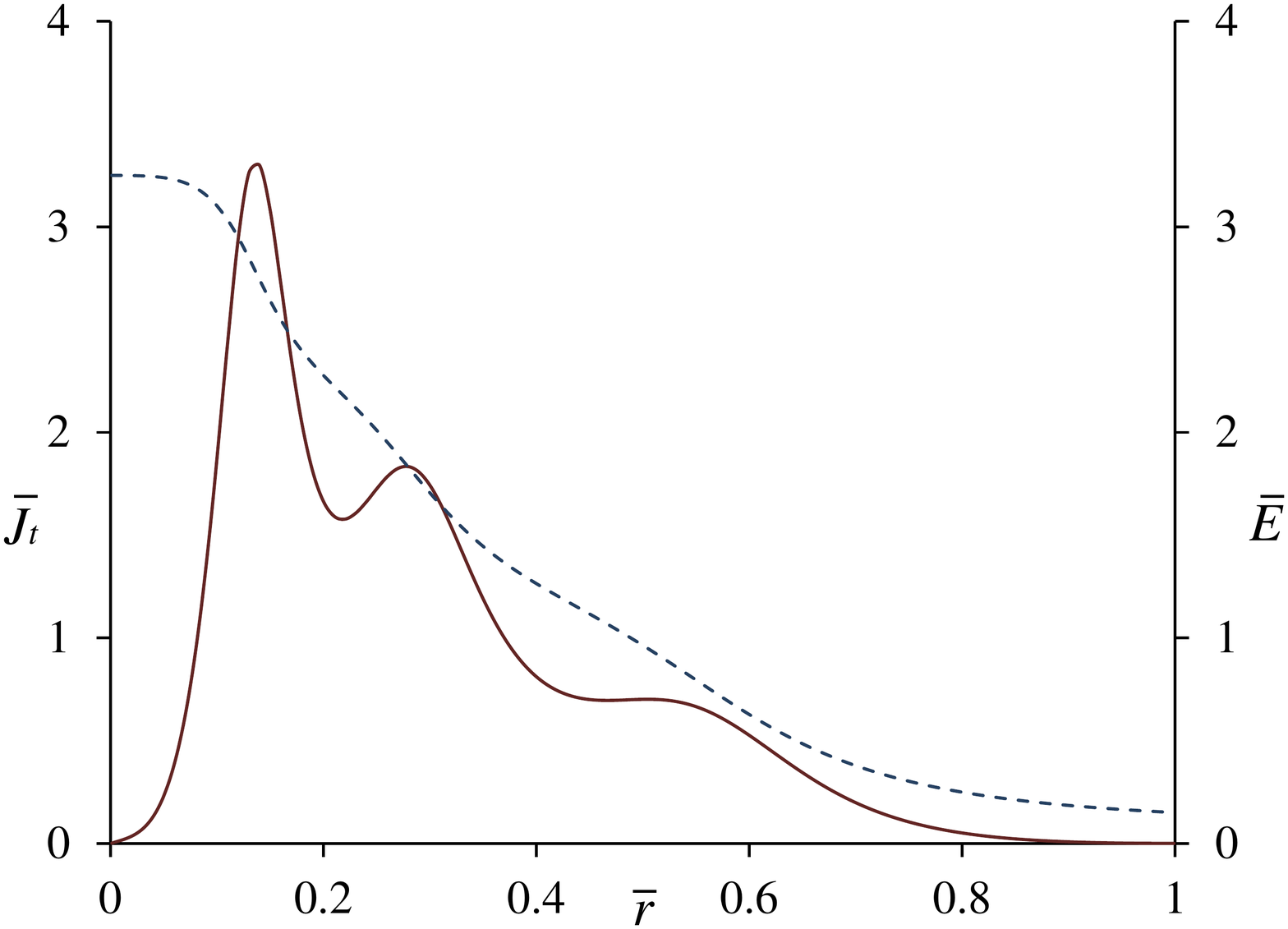}}
  \caption{Typical electron star configurations in the black hole. The solid lines show the charge density, $\bar{J}_t=fr_s\partial_r \phi/{2\pi}$ as a function of radial position $\bar{r}=r/r_h$. The dashed lines show the normalised electric field, $\bar{E}=r_s^2F_{rt}/\pi $.  The left-hand graph has filling fraction $\phi_0/2\pi = 2$ and is plotted $m_R^2L^2 = 4$ and ${r_h^2}/{r_s^2} = 5$. Note that only a single peak is visible; this is typical for low values of the mass.    The right-hand graph is plotted with $m_R^2L^2=16$ with  ${r_h^2}/{r_s^2} = {10}$ with a fractional filling number  $\phi_0/2\pi=3.25$. Note that this results in a residual  electric field at the horizon of the black hole.}
\label{bhfig}
\end{figure}
\noindent
The first term is due to the mass of the fermion. The second term arises after integrating out the gauge field and can be thought of as capturing the electrostatic energy between fermions.

\para
We again solve \eqn{does1} numerically. Results for typical values of the parameters\footnote{These plots are made in the regime $r_s\ll r_h$ or, equivalently, $e^3 L \gg \kappa$. This ensures that the screening, and hence kink formation, takes place before we hit the horizon.  In the opposite regime, $r_s\gg r_h$, the electric field is largely unscreened for much of its profile, although some screening does take place near the horizon.} are plotted in Figure \ref{bhfig}: solid lines denote the local bulk charge density, dashed lines the electric field. These plots look very similar to those in Figure \ref{hardfig} because most of the kinking is taking place in the AdS$_4$ region of the geometry. However, there are differences. Most importantly, the allowed charge density is no longer restricted to integer filling fraction. Solutions with fractionally filled Landau levels happily exist and we have presented one in the right-hand graph. Such solutions are typically (but not always) accompanied by a residual electric field  at the horizon, showing the existence of fractionalised charge. We will devote Section \ref{fcsec} to a deeper analysis of the relationship between fractional filling and fractionalised charge. (Just because they're both called named after fractions doesn't mean they're the same thing!)



\begin{figure}[!h]
  \begin{center}
    \includegraphics[trim = 0.5in 0.5in 0.2in 0.7in, width=3in]{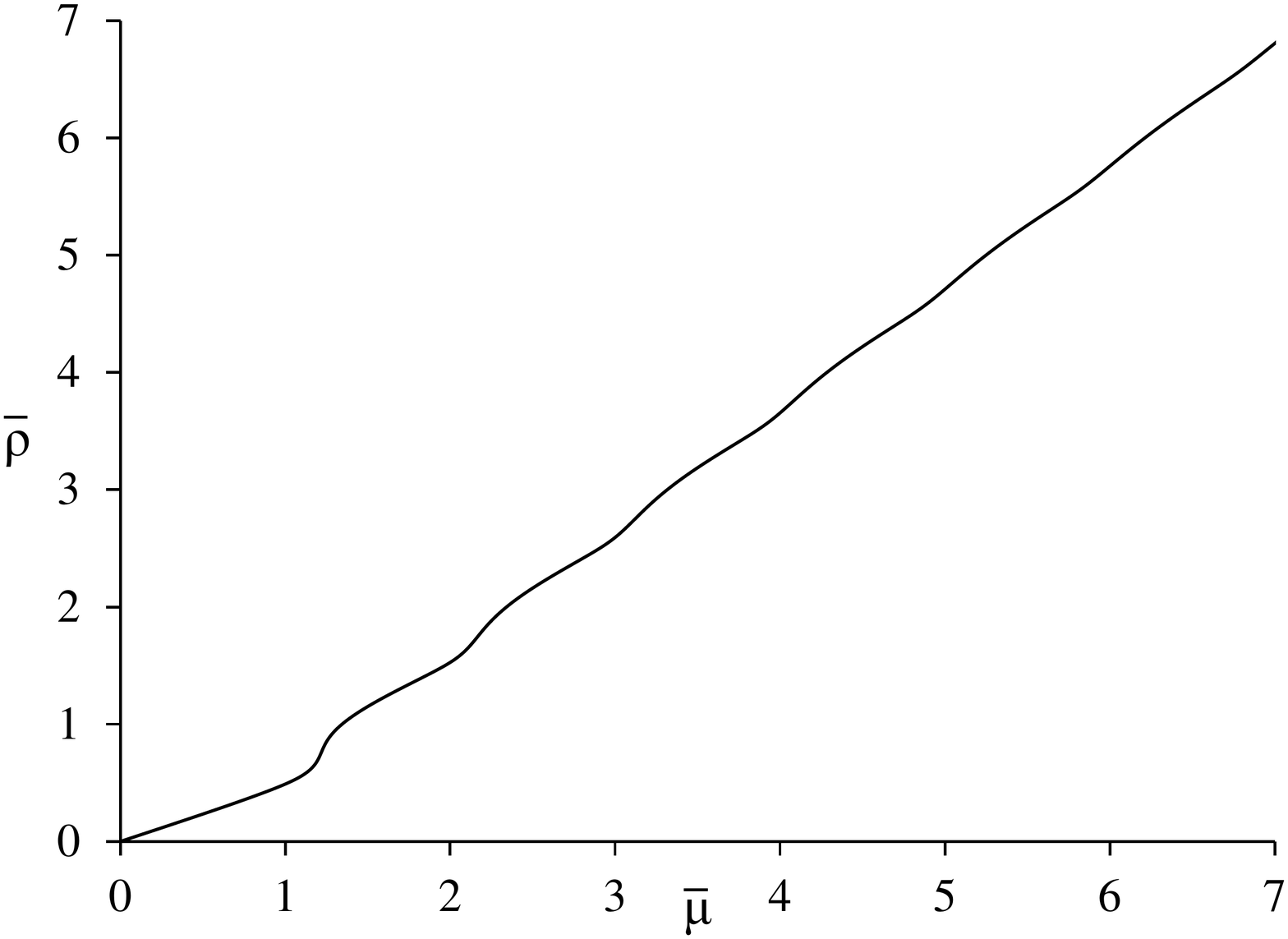}
  \end{center}
  \caption{A graph of how the normalised charge density in the boundary theory, $\bar{\rho} = \rho r_s^2/\pi = {\phi_0/}{2\pi}$, varies with the chemical potential, $\bar{\mu}=\mu r_s/\pi$. 
 Note that the vertical axis corresponds to filling fraction. The normalization is such that in this, and all other plots, the massless fermion gives rise to a straight line of unit slope.
  This graph was made with $m_R^2L^2 = 6$ and  ${r_h^2}/{r_s^2} =8$.  }
  \label{rhofig}
\end{figure}

\subsection{Charge Jumps}\label{platsec}

As we will now see, the presence of the kinks in the bulk shows up in a number of physical  properties of the boundary field theory. Perhaps the simplest  physical observable is the boundary charge density, $\rho$. This can be plotted as a function of the chemical potential $\mu$, given by integrating the electric field between the boundary and horizon
\be
\mu = \frac{e^2 B}{4\pi^2} \int_0^{r_h} dr (\phi_0-\phi)
\ee
The relationship between $\rho$ and $\mu$ in our system is somewhat intricate and changes qualitatively as various parameters are varied. In this section and the next,  we take some time to describe this.

\para

For massless fermions, the relationship between the charge density and chemical potential is linear. As we turn on the mass $m$ of the fermion and, correspondingly, the sine-Gordon potential for the bosonized scalar,  kinks form in the bulk. Their effect is to induce wiggles in the relationship between $\rho$ and $\mu$. These wiggles occur close to integer filling fractions and are largest for small $\rho$, dying out as the charge density is increased. A typical example is shown in Figure \ref{rhofig}.

\begin{figure}[!h]
  \begin{center}
    \includegraphics[trim = 0.5in 0.5in 0.2in 0.7in, width=3in]{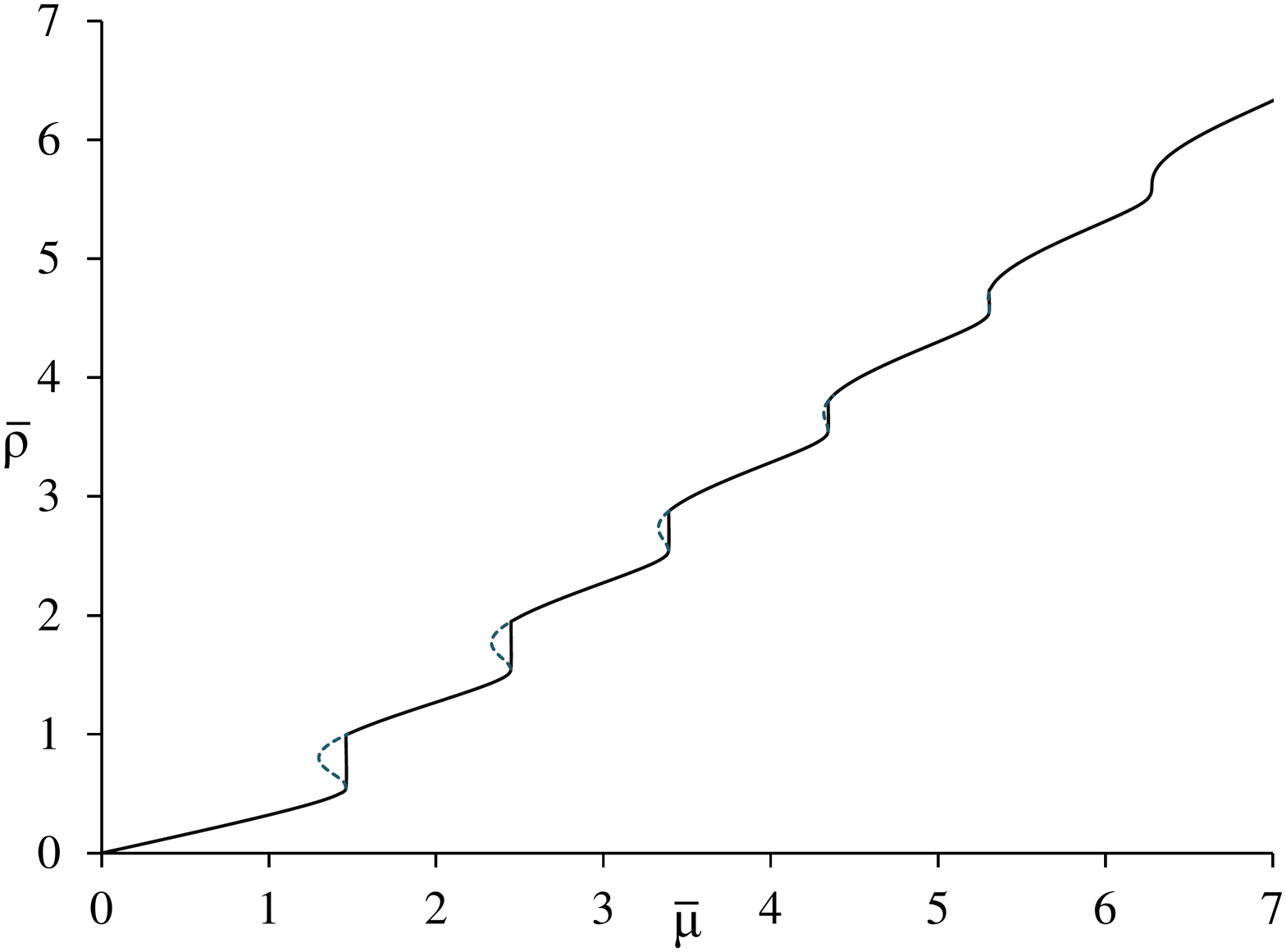}
  \end{center}
  \caption{The total charge density in the boundary theory, $\bar{\rho}=\rho r_s^2/\pi = {\phi_0/}{2\pi}$. This graph was made with parameters $m_R^2L^2=12$ and $r_h^2/r_s^2=20$. Note the discontinuous jumps in the first five bands.}.
  \label{upfig}
\end{figure}

\para
It is worth stressing the obvious: the relationship between $\rho$ and $\mu$ in this regime is continuous. We can contrast this to  the situations for free fermions in a magnetic field which would result in step-like behaviour every time  the chemical potential crosses the energy of each Landau level. As we have already noted, in our strongly interacting theory, the energy levels of the bands are broadened and the steps are smoothened out, with only ripples surviving.

\para
However, as the mass $m_R^2L^2$ is increased further, the wiggles in the $\rho$ vs $\mu$ plot become more pronounced. Eventually, they grow so large that the graph $\rho(\mu)$ is no longer single valued. A characteristic example\footnote{This graph is created by working backwards, fixing $\rho$ and then determining $\mu$. The function $\mu(\rho)$ is single-valued, but is not monotonic.}  is shown in Figure \ref{upfig}. For values of $\mu$ which allow solutions with two different charge densities $\rho$, one must compute the free energies to determine which wins. We will discuss the free energy of the electron stars in Section \ref{qosec} --see equation \eqn{ham} -- in the context of quantum oscillations. In the present case, the result is simple: the solution with the smaller free energy is that with the smaller value of $\rho$. This results in a discontinuous jump in the charge density as the chemical potential is increased.  (A similar  quantum Hall-like step was seen for lowest Landau level in the  D3-D7 probe brane system \cite{bergman1,bergman2}).

\para
The discontinuities in the total charge density $\rho$ occur only for the lowest bands. As the mass is increased (all other parameters remaining fixed), more and more bands suffer the discontinuity.

\subsection{Fractionalised and Cohesive Charge}\label{fcsec}

Above, we have seen that there are discontinuities in the total charge density only if the mass is suitably large. However, even in the case that $\rho$ appears smooth -- as depicted  in Figure \ref{rhofig} -- if we look more closely we can see that the familiar Landau level steps are not as far away as one might imagine.  As we reviewed at the start of this section, charge density in the holographic framework comes in two forms: cohesive charge, carried by bulk fields, and fractionalised charge, hidden behind the horizon. Let us see how the charge density $\rho$ is split  between these two options. 

\para
The amount of fractionalised charge is determined by the electric field at the horizon. Using \eqn{doing1}, we can write this as
\be
\rho_{\rm{frac}} = \frac {B} {4\pi^2} (\phi_0-\phi(r_h))
\ee
From the discussion in the previous section, we know that $\phi(r_h)$ necessarily lies at an extremum of the effective potential \eqn{vir}.

\begin{figure}[!h]
  \resizebox{80mm}{!}{\includegraphics{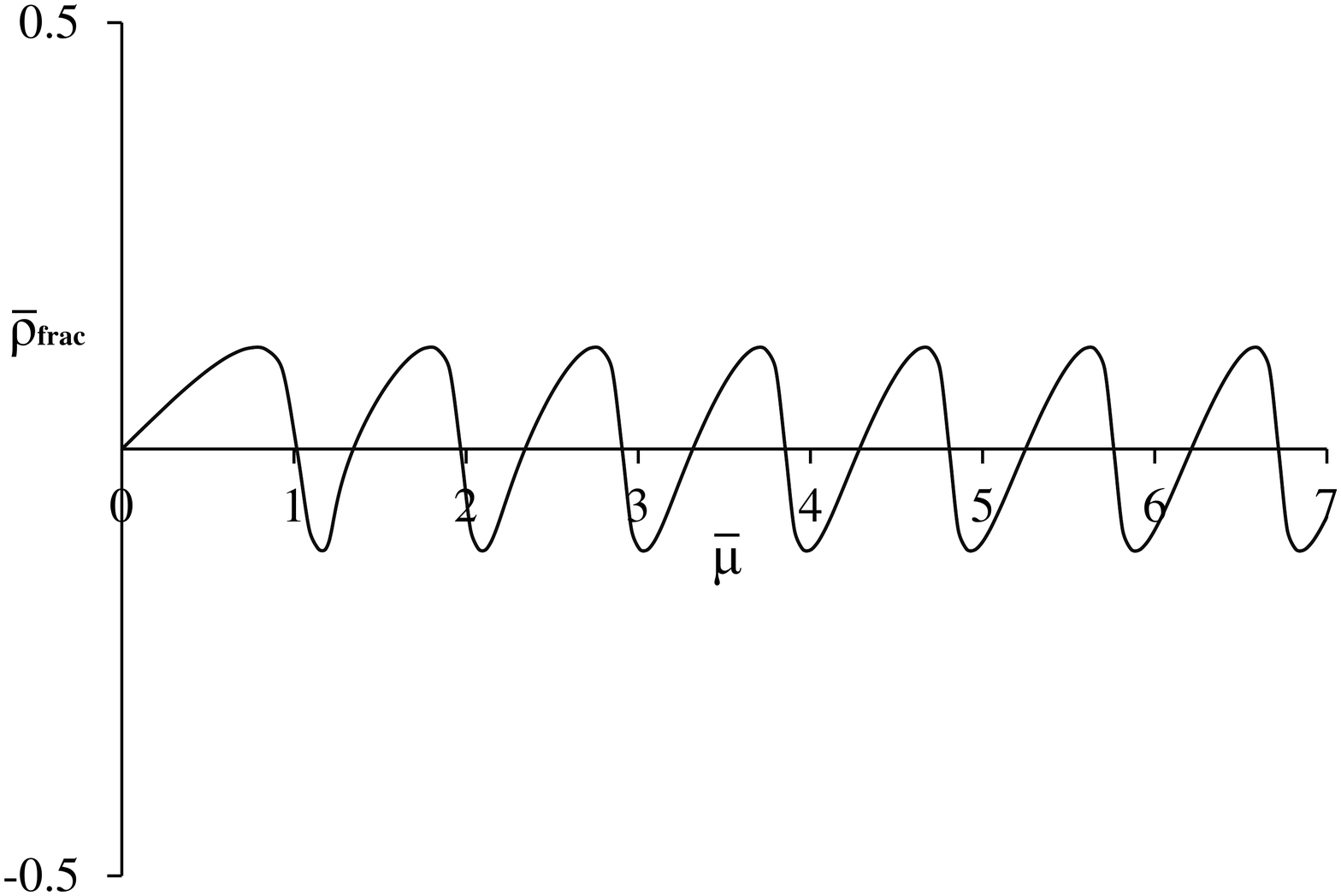}}
\resizebox{80mm}{!}{\includegraphics{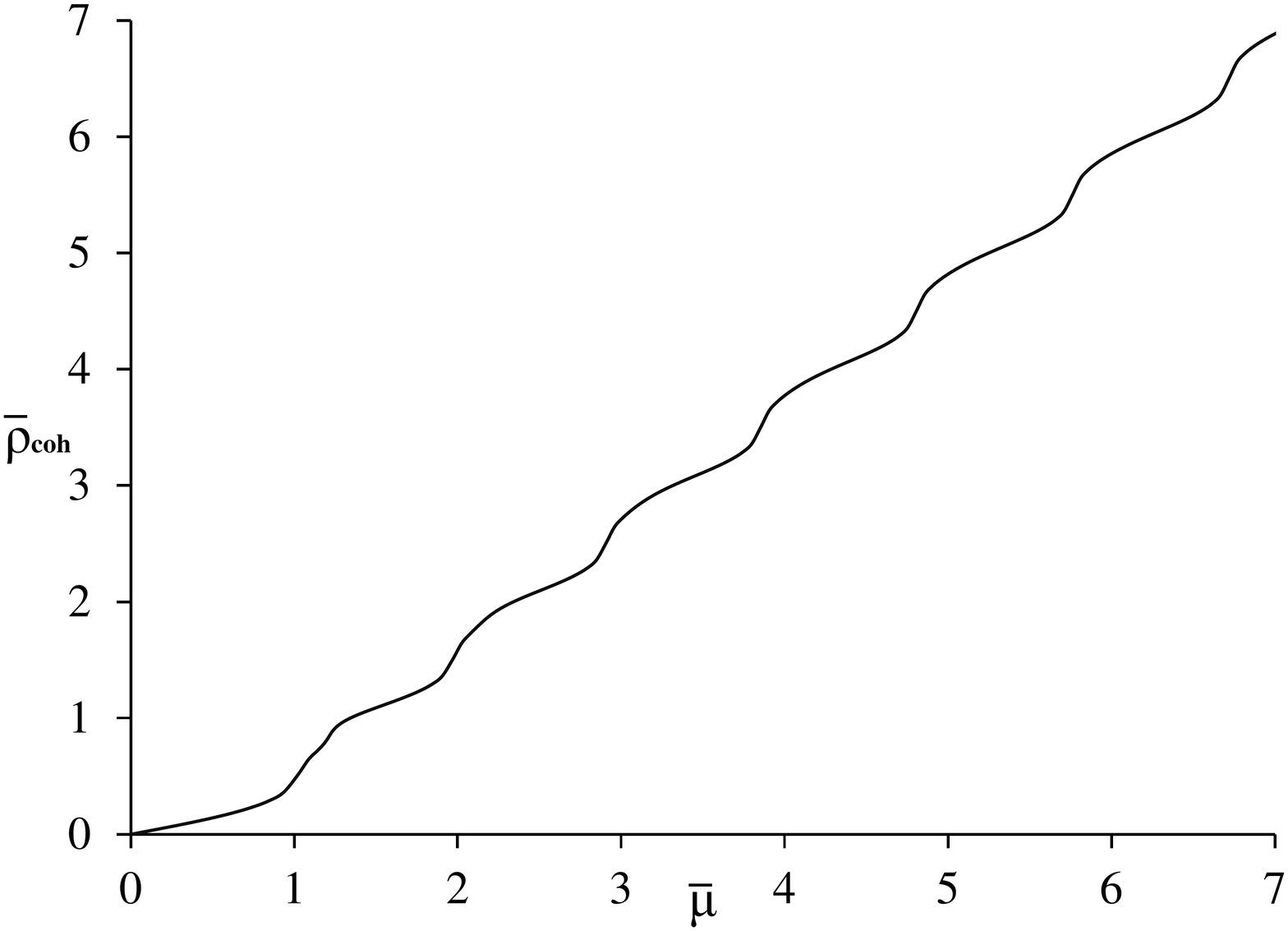}}
  \caption{When $m^2_RL^2 < r_h^2/r_s^2$, both the fractionalised charge density (on the left) and the cohesive charge density (on the right) are continuous. These plots have the same parameters as Figure 3: $m_R^2L^2 = 6$ and $r_h^2/r_s^2=8$.}
\label{fracfig}
\end{figure}

\begin{figure}[!h]
  \resizebox{80mm}{!}{\includegraphics{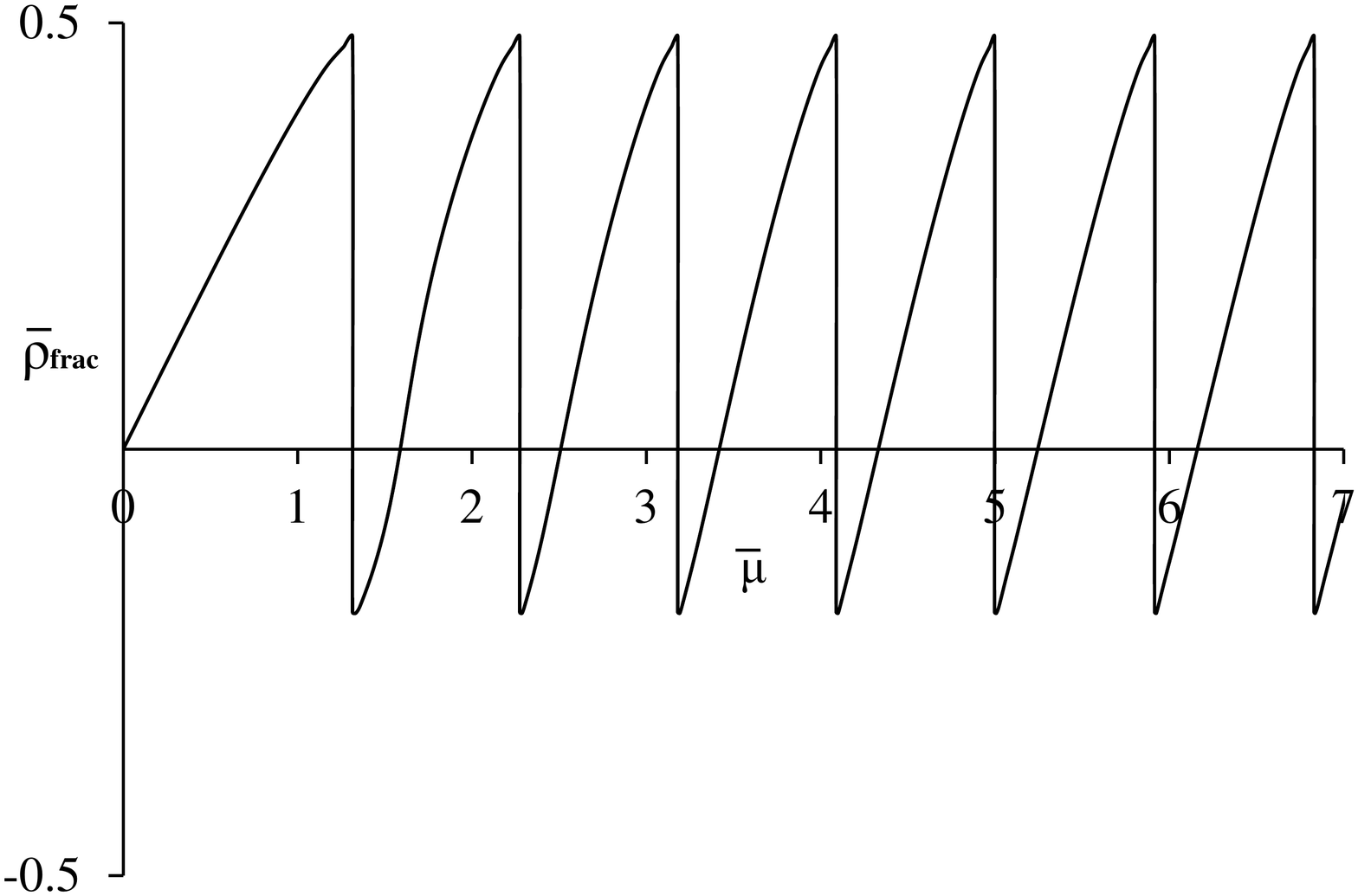}}
\resizebox{80mm}{!}{\includegraphics{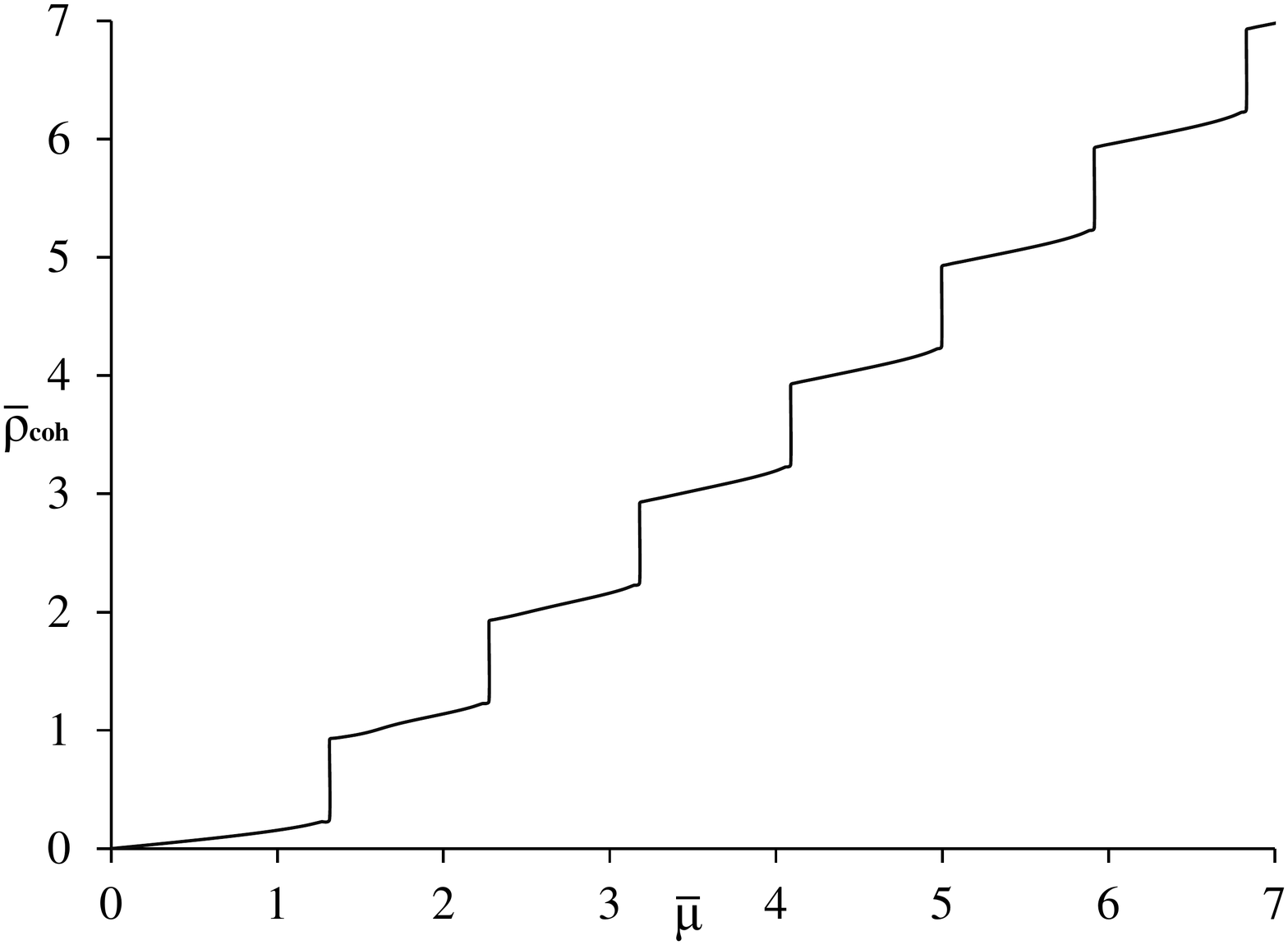}}
  \caption{When $m^2_RL^2 > r_h^2/r_s^2$, both the fractionalised charge density (on the left) and the cohesive charge density (on the right) are show discontinuities. Here we have taken $m_R^2L^2 = 12$ and $r_h^2/r_s^2=4$. When combined the two discontinuities  cancel out to form the smooth total charge density that does not look qualitatively different from Figure 3.}
\label{jumpfig}
\end{figure}

\para
It is simple to extract the amount of fractionalized charge  $\rho_{\rm frac}$ from our numerical solutions. We find that as the chemical potential is increased, the fractionalised charge density oscillates about zero. (Examples can be seen in Figures \ref{fracfig} and \ref{jumpfig}). Using the considerations above, it  is straightforward to check that in all cases the amplitude of oscillation is bounded by
\be |\rho_{\rm frac}| \leq \frac{m_R^2L^2}{2e^2r_h^2}\label{upperbound}\ee
Notice in particular that, in the case of massless fermions, all the charge is carried by the electron star; none by the horizon.

\para
To understand the fractionalised charge density in more detail, we need to look at the extrema of the potential \eqn{vir}. These depend 
 on the relative size of  $m_R^2L^2$ vs $r_h^2/r_s^2$. As we now show, this gives rise to a qualitative difference  in the behvaiour of the charge density.

\subsubsection*{{When $\rho_{\rm frac}$ Changes Continuously...}}

We start by considering $m_R^2L^2<  r_h^2 / r_s^2$. In this regime,  $V_{IR}(\phi)$ has a unique extremum for any value of filling fraction $\phi_0/2\pi$. This extremum is a global minimum and regularity at the boundary means that $\phi(r_h)$ must sit in this minimum. With no ambiguity in the solution, the fractionalised charge is a continuous function of $\mu$. The fractionalised charge that contributes to the total charge density shown in Figure \ref{rhofig} is shown in the left-hand graph of Figure \ref{fracfig}. Note that the fractionalised charge can be both positive and negative.

\para
Of course, if the fractionalised charge oscillates and the total charge is monotonically increasing, the remainder must be made up in cohesive charge. This is shown in the right-hand graph in Figure \ref{fracfig}.

\para
Notice that the fractionalised charge density vanishes for certain values of the chemical potential. Translating this into filling fraction, we find that $\rho_{\rm frac}=0$ whenever a Landau level is fully filled or half filled. This can be understood by looking again at the effective potential \eqn{vir}. The charge at the horizon vanishes whenever $\phi=\phi_0$ is the minimum of this potential. This  holds for $\phi=n\pi$ with $n\in {\bf Z}$. Integer filing corresponds to even $n$; half-integer to odd $n$. At half-integer filling, $\phi$ sits at a maximum of the $(1-\cos\phi)$ potential, giving rise to an extra half kink in the bulk. But, by virtue of the condition $m_R^2L^2<  r_h^2 / r_s^2$, this  maximum of $(1-\cos\phi)$ becomes the minimum of the  full effective potential \eqn{vir} which includes the quadratic electrostatic term. 

\para
The fact that $\rho_{\rm frac}=0$ at integer filling is unsurprising. The fact that $\rho_{\rm frac}=0$ also at half-integer filling in this regime is surprising. We do not understand this from the perspective of the   boundary theory.

\subsubsection*{{When $\rho_{\rm frac}$ jumps...}}

As we increase the mass past the  critical value, $m^2_RL^2>r_h^2/r_s^2$, things get more interesting.  In this regime, there exist some values of filling fraction $\phi_0/2\pi$ for which there are \emph{multiple} extrema of $V_{IR}(\phi)$. A full study of the solutions is a little complicated and is described  in Appendix \ref{appb}. Despite multiple minima, only one value of $\phi(r_h)$ gives rise to a stable field configuration at any filling fraction. This preferred $\phi(r_h)$ is always a local minimum of $V_{IR}(\phi)$,
 but not necessarily a global minimum. Furthermore, as we vary the chemical potential, there are values at which the preferred minimum jumps discontinuously. At these points, the fractionalised charge also jumps. A typical example is shown  in the left-hand graph of Figure \ref{jumpfig}.

 \para
 The discontinuity also arises in the cohesive charge density. A plot is shown in the right-hand side of Figure \ref{jumpfig} and exhibits the characteristic step features expected of Landau levels. (Although these are really different bands, all of which lie in the lowest Landau level).  These steps continue indefinitely for higher bands. Note, however, that the plateaux are not precisely horizontal; the Landau levels in the strongly interacting boundary theory are not precisely flat bands. 

\para
We stress that the step-like behaviour seen in the cohesive charge is not obviously related to the discontinuities appearing in the total charge that we exhibited in the previous section. In particular, the jumps in the fractionalised and cohesive charge shown in Figure \ref{jumpfig} occur both in the regime where the total charge is monotonic (as shown in Figure \ref{rhofig}) and also in the regime where the total charge itself exhibits a finite number of jumps (as shown in Figure \ref{upfig}). 

\begin{figure}[!h]
  \begin{center}
    \includegraphics[trim = 0.5in 0.5in 0.2in 0.7in, width=3in]{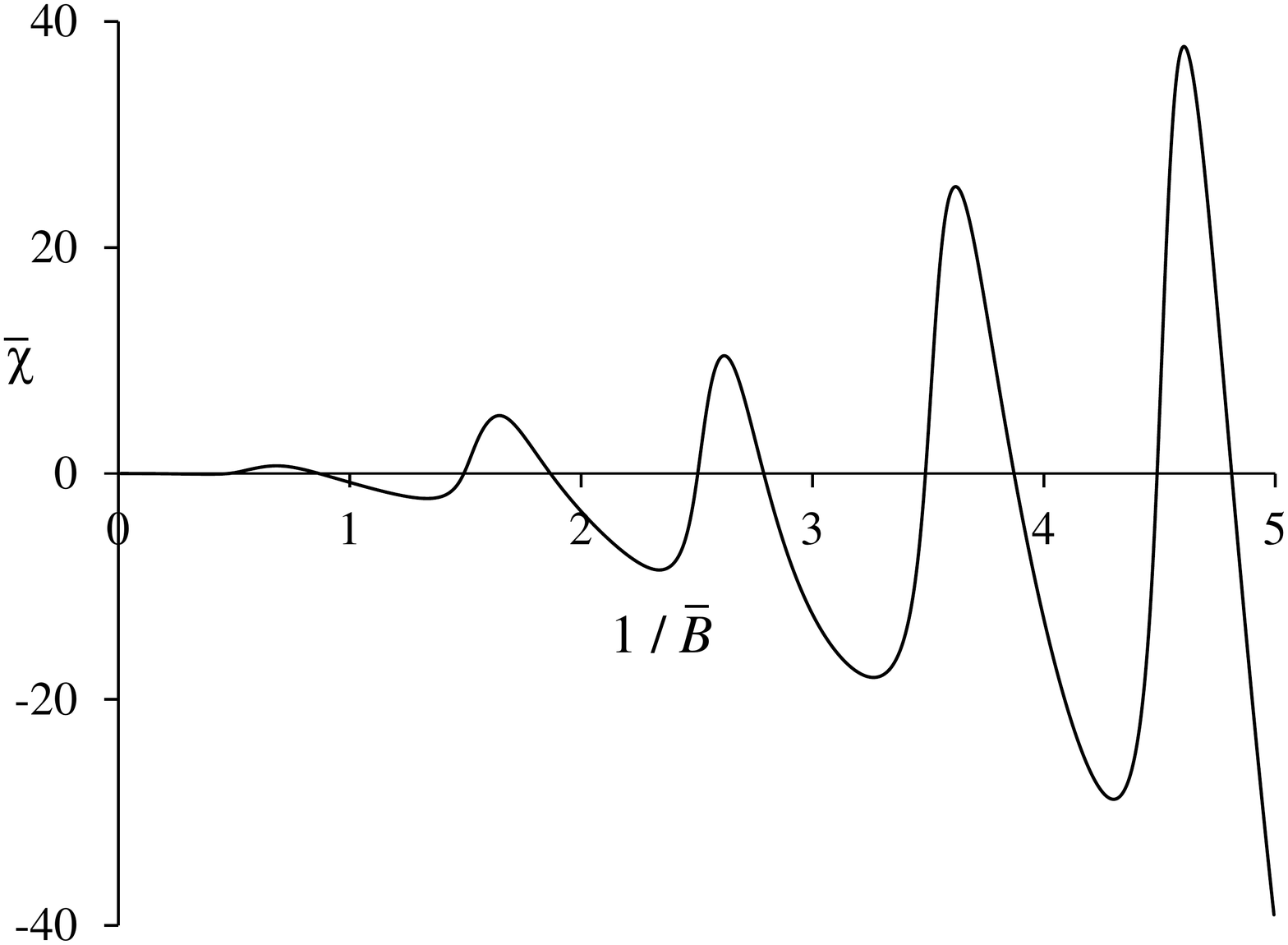}
  \end{center}
  \caption{The dimensionless magnetic susceptibility, $\bar{\chi}=  (6e^2L^2\rho^2/\kappa^2)^{1/4}\, \chi$ plotted against the filling fraction, $\phi_0/2\pi$. The parameters chosen for this plot are $m_R^2L^2 = 12$, ${r_h^2}/{r_s^2} = 20$.}
  \label{dHvA}
\end{figure}

\para
Of course, for free fermions the lowest Landau levels are the domain of quantum Hall physics. In this paper, we have restricted attention to equilibrium properties of the system and have not attempted to compute transport. Needless to say, it would be of great interest to do so, although the techniques of bosonization do not obviously lend themselves to the task. Nonetheless, plots such as Figure \ref{jumpfig} show that one will find oscillatory or step-like behaviour for any process in which the fractionalised and cohesive charges contribute differently. In the next section, we present a simple example.

\subsection{Quantum Oscillations}\label{qosec}

In this section, we  wish to compute how the magnetic susceptibility  varies as a function of magnetic field $B$ at fixed charge density $\rho$. As we increase $B$, the occupancy of each Landau level increases and, correspondingly, the filling fraction decreases.. Each time the number of filled bands changes, the magnetic susceptibility oscillates. These are the famous de Haas van Alphen oscillations, although in our case we are filling the lowest Landau levels of different bands as opposed to higher Landau levels of the same band.  From \eqn{likethis}, the period of oscillation is 
\be 
\Delta\left(\frac 1 B \right) = \frac{1}{2\pi \rho}
\ee
For three-dimensional metals, this is often written as $\Delta(1/B) = 2\pi e/A_F$ where $A_F$ is the extremum area of the Fermi surface.

\para
To compute the magnetic susceptibility, we first need to evaluate the free energy of the bulk fermions and electric field. Because we are working at zero temperature, this simply means the energy. Integrating out the electric field from \eqn{bosoned} yields an effective Hamiltonian for $\phi$,
\be
H_{\rm{eff}} = \frac{BA}{2\pi} \int_0^{r_h}dr\ \left( \frac{1}{8\pi} f(r) (\partial_{r} \phi)^2 +\frac{m_R^2L^2}{4\pi r^2}(1 - \cos \phi)+\frac 1 {8\pi r_s^2} (\phi - \phi_0)^2\right) 
\label{ham}\ee
Note that this is closely related to the effective potential defined in \eqn{vir}. 
The magnetic susceptibility is then given by evaluating the on-shell Hamiltonian and taking the second derivative,
\be \chi \equiv -\frac{1}{A}\frac{\partial^2 H_{\rm eff}}{\partial B^2}\nn\ee
The result is plotted in Figure \ref{dHvA} and clearly exhibits quantum oscillations. 
This particular plot was made in a regime where both the total charge density and the fractionalised charge density vary smoothly (as in Figures \ref{rhofig} and \ref{fracfig}). However, the figure looks qualitatively the same for the other regimes. Notice, in particular, that we did not need to turn on a temperature to smooth out the oscillations; quantum effects have already achieved this for us\footnote{There is a caveat to this statement. Our quantum oscillations were computed at fixed $\rho$ rather than fixed $\mu$. If we chose to work in the grand canonical ensemble, then for suitably large masses we find a first order phase transition as the total charge density jumps as shown in Figure \ref{upfig}. In this case, it seems that finite $T$ is necessary to smooth out the quantum oscillations. Finite temperature was also found to be necessary to compute quantum oscillations  in the original electron star story \cite{sean2}.}.

\para
The free energy computed above  does not contain the contribution from the gravitational background. This contains an overall contribution from the \RN\ black hole which, at zero temperature scales as $\chi \sim 1/\sqrt{B}$, providing an overall envelope for the contribution above. There will furthermore be an oscillatory component due the horizon moving backwards and forwards as the fractionalised charge waxes and wanes. However, this contribution depends on a new parameter,  $\kappa$, and cannot remove the oscillations from the electron star computed above. 

\para
Quantum oscillations were seen previously in the probe approximation \cite{koslif1,koslif2}, where they are a sub-leading $1/N$ effect, and also in electron stars computed in the fluid approximation \cite{sean2}, where the standard Kosevich-Lifshitz formula for their amplitude was reproduced. It would be interesting to understand the amplitude and temperature dependence of the oscillations in the present case. While we have not made analytic progress with this question, numerics suggest that the increase in the amplitude of oscillations is fairly well approximated by the $\chi\sim 1/B^2$ behaviour of the Kosevich-Lifshitz formula.

\section*{Acknowledgements}

We would like to thank Aleksey Cherman, Nick Dorey and Daniele Dorigoni for useful discussions on the art of bosonization and Sean Hartnoll for comments on the manuscript. SB is supported by the Israel Science Foundation Center of Excellence and by the Lady Davies fellowship. MB, DT and KW are supported by STFC and by the ERC STG grant 279943, ``Strongly Coupled Systems".

\appendix
\section{Appendix: Einstein's Equations and the Stress Tensor}\label{appa}

The purpose of this appendix is to construct the 4d stress tensor in bosonized form. As we shall see, the requirement that the stress tensor is conserved puts severe constraints on the method of bosonization in curved space and, in particular, the choice of regularization scale $\Lambda$. 

\para
With the stress tensor in hand, we will write the Einstein equations for translationally invariant states in bosonized form. Although we do not solve these equations, we will check that in the regime of interest the gravitational solution is dominated by the magnetic field.


\para
The Einstein equations resulting from the action \eqn{action} take the form
\be
R_{\mu\nu} - \tfrac 1 2 R g_{\mu\nu} - \frac{3}{L^2} g_{\mu\nu} = \kappa^2 T_{\mu\nu}.\nn
\ee
In the conformal $(\tilde r, t, x, y)$ coordinates, with metric ansatz \eqn{metric2}, the various components of the Einstein tensor are
\be
R_{tt} - \tfrac 1 2 R g_{tt} - \frac{3}{L^2}g_{tt} &=& - \Sigma^{-2} (\partial_{\tilde r} \Sigma)^2 - 2\Sigma^{-1} \partial_r^2 \Sigma + 2\Omega^{-1} \Sigma^{-1} \partial_r \Omega \partial_{\tilde r} \Sigma + \frac{3\Omega^2}{L^2} \nn \\
R_{\tilde r \tilde r} - \tfrac 1 2 R g_{\tilde r \tilde r} - \frac{3}{L^2}g_{\tilde r \tilde r} &= & \Sigma^{-2} (\partial_{\tilde r} \Sigma)^2 + 2\Omega^{-1} \Sigma^{-1} \partial_{\tilde r} \Omega \partial_{\tilde r} \Sigma -  \frac{3\Omega^2}{L^2} \nn \\
R_{xx}-\tfrac 1 2 R g_{xx} - \frac{3}{L^2} g_{xx} &=& \Omega^{-2} \Sigma \partial_{\tilde r}^2 \Sigma + \Omega^{-3} \Sigma^2 \partial_{\tilde r}^2 \Omega - \Omega^{-4} \Sigma^2 (\partial_{\tilde r} \Omega)^2 -  \frac{3\Sigma^2}{L^2}\nn
\ee
The stress tensor contains contributions from both the gauge fields and fermions
\be T_{\mu\nu} = T_{\mu\nu}^{gauge} + T_{\mu\nu}^{\rm fermion}
\nn\ee
The contribution from the Maxwell term can be derived by varying the action with respect to the metric
\be  T_{\mu\nu}^{gauge}= \frac 1 {e^2} (F_{\mu\rho}F_\nu^{\rm{  }\rho} - \tfrac 1 4 g_{\mu\nu} F_{\rho\sigma}F^{\rho\sigma}) \nn\ee
Meanwhile, the stress tensor for fermions  in curved space can be derived by varying the action with respect to the vierbein. (See, for example, \cite{freedman}). The result is
\be T_{\mu\nu}^{\rm fermion} = \left(\tfrac 1 4 \bar\psi e_{{a} \mu} \left( \partial_\nu + \tfrac 1 8 \omega_{\nu, bc} [\Gamma^{b}, \Gamma^{c} ] - iA_\nu \right)\Gamma^{a}\psi + \rm{h.c.} \right) + (\mu \leftrightarrow \nu).\nn
\ee
Since we are interested only in states that enjoy translational invariance in the $x-y$ plane, we can write
\be
T_{\mu\nu} = \frac 1 A \int dxdy \ T_{\mu\nu},
\nn\ee
Substituting in the mode expansion \eqn{decomp}, and performing the integrals over $x$ and $y$, we arrive at expressions for the stress tensor in terms of the Landau level fermions
\begin{eqnarray*}
T_{tt}^{\rm fermion} &=& - \frac 1{2A\Sigma^2}\sum_k \left( \sum_{n=0}^\infty i\bar\xi_{-\,n,k} \gamma^t (\partial_t - iA_t)\xi_{-\,n,k} + \sum_{n=1}^\infty i\bar\xi_{+\,n,k} \gamma^t (\partial_t - iA_t) \xi_{+\,n,k} + \rm{h.c.}  \right) \nn \\
T_{\tilde r \tilde r}^{\rm fermion} &=& \frac{1}{2A\Sigma^2}\sum_k \left( \sum_{n=0}^\infty i\bar\xi_{-\,n,k} \gamma^r \partial_{\tilde r} \xi_{-\,n,k} + \sum_{n=1}^\infty i\bar\xi_{+\,n,k} \gamma^r \partial_{\tilde r} \xi_{+\,n,k} + \rm{h.c.} \right)  \nn \\
T_{\tilde rt}^{\rm fermion} &=& \frac 1{4A\Sigma^2}\sum_k  \left( \sum_{n=0}^\infty (i\bar\xi_{-\,n,k} \gamma^r (\partial_t - iA_t) \xi_{-\,n,k} - i\bar\xi_{-\,n,k} \gamma^t \partial_{\tilde r} \xi_{-\,n,k}) \right. \nn \\
&& \qquad \qquad \qquad \qquad  \quad \left. + \sum_{n=1}^\infty (i\bar\xi_{+\,n,k} \gamma^r (\partial_t - iA_t) \xi_{+\,n,k} -i\bar\xi_{+\,n,k} \gamma^t \partial_{\tilde r} \xi_{+\,n,k}) + \rm{h.c.}  \right) \nn \\
\end{eqnarray*}
As is clear from these expressions, each of these components receives contributions from all Landau levels. However, when we come to compute the $T_{xx}^{ \rm fermion}$ and $T_{yy}^{ \rm fermion}$ components of the stress tensor, only the higher Landau levels contribute, 
\be T_{xx}^{ \rm fermion} = T_{yy}^{ \rm fermion} =  \frac 1 {2A\Sigma\Omega}\sum_k  \left( \sum_{n=1}^\infty i \sqrt{2Bn}\bar\xi_{n+} \gamma^3 \xi_{n-} + \rm{h.c.} \right) 
\nn\ee
This is another manifestation of the fact that fermions in the lowest Landau level undergo an effective dimensional reduction  and are, therefore, susceptible to bosonization. Since we will excite only fermions in the lowest Landau level, we are interested in solutions with $T^{\rm fermion}_{xx} = T^{\rm fermion}_{yy}=0$. Note that this automatically means that we are working in a regime that is far from the isotropic stress energy tensor assumed in Thomas-Fermi approximations to the electron star. 

\para
Restricting now to fermions in the lowest Landau level, we would like to construct expressions for $T_{tt}^{\rm fermion}$, $T^{\rm fermion}_{\tilde{r}\tilde{r}}$ and $T_{\tilde{r} t}^{\rm fermion}$ in terms of the bosonized field $\phi$. In fact, requiring a consistent stress tensor provides very non-trivial constraints on the choice  of regularization scale $\Lambda$. Specifically, we will ask that our stress tensor is gauge invariant, symmetric and, most importantly, conserved when the equations of motion are obeyed
\be \nabla^\mu T_{\mu\nu}=0\nn\ee
%
%
%
These criteria are only satisfied if we pick $\Lambda \sim m$. The resulting expression for the stress tensor for translationally invariant and time invariant states is
\be
T_{tt} &=& \frac{1}{\Sigma^2} \left( \frac {B} {2\pi}\right) \left( \frac 1 {8\pi} (\partial_{\tilde r} \phi)^2 +\frac{m_R^2L^2 \Omega^2}{4\pi} (1-\cos \phi) \right) + \frac 1 {2e^2} \left( \frac 1 {\Omega^2} (\partial_{\tilde r} A_t)^2 + \frac{\Omega^2}{\Sigma^4}B^2 \right) \nn \\
T_{\tilde r \tilde r} &=& \frac 1 {\Sigma^2} \left( \frac {B} {2\pi}\right) \left( \frac 1 {8\pi} (\partial_{\tilde r} \phi)^2 -\frac{m_R^2L^2 \Omega^2}{4\pi} (1-\cos \phi) \right) - \frac 1 {2e^2} \left( \frac 1 {\Omega^2} (\partial_{\tilde r} A_t)^2 + \frac{\Omega^2}{\Sigma^4}B^2 \right) \nn \ee
Furthermore, the off-diagonal component vanishes for time-independent states, $T_{\tilde{r}t}=0$, while, as described above, the pressure in the $x$ and $y$ directions gets contributions only from the gauge field, 
\be
T_{xx} &=& T_{yy} =  \frac{1}{2e^2} \left( \frac {\Sigma^2} {\Omega^4} (\partial_{\tilde r} A_t)^2 + \frac 1 {\Sigma^2}B^2 \right)
\nn\ee
%
As a further check of this result, note that when the $\tilde{r}-t$ plane is flat, the stress tensor should coincide with the Noether currents arising arising from translational invariance: it does.  

\para
Notice that altogether we have five equations of motion, but only four degrees of freedom. As usual, there is no inconsistency because the Bianchi identity and the conservation of the stress tensor imply that one of the equations is redundant. 

\subsubsection*{Einstein Equations}

Using the bosonized stress tensor, we can now derive the coupled Einstein-Maxwell-Dirac equations in bosonized form, capturing the gravitational backreaction of fermions in the lowest Landau level. The resulting equations are simplest if we revert to the $r$ coordinates of \eqn{metric} and parameterize the metric as
\be
ds^2 = \frac{L^2}{r^2} \left(- e^{2C(r)}D(r) dt^2 + \frac 1 {D(r)} dr^2 +dx^2+dy^2 \right)
\nn\ee
Gauss' law can be integrated trivially to give
\be
\partial_r A_t = -\frac{e^2 B e^C }{4\pi^2} (\phi - \phi_0),
\nn\ee
and the other fields obey a simple coupled system of ordinary differential equations.
\be
\partial_r C &=& -\frac{\kappa^2}{16\pi^2} \, \frac{Br^3}{L^2} (\partial_r \phi)^2 \nn \\
\partial_r D  & =& \frac{3(D-1)}{r}+  \frac{\kappa^2Br^3}{16\pi^2L^2}\left[ D (\partial_r \phi)^2 +\frac{2m_R^2L^2}{r^2}  (1-\cos \phi ) + \frac{8\pi^2 B}{e^2}\left( 1 + \frac{e^4}{16\pi^4} (\phi - \phi_0)^2\right) \right] \nn \\
\partial_r^2 \phi & = & -\left( \frac{\partial_r D} D + \partial_r C \right) \partial_r\phi +\frac{m_R^2L^2}{ r^2 D} \sin \phi + \frac{ e^2 B}{2\pi^2 D} (\phi - \phi_0)
\nn\ee

\subsubsection*{Background Check}

We now use the results above to determine when we can neglect the backreaction of the fermions and electric field on the metric.  This holds when the dominant contribution to $T_{tt}$ is provided by the  magnetic field. The relevant equation is
\be
T_{tt} = \frac{Br^2f(r)}{2\pi}  \left( \frac 1 {8\pi} f(r) (\partial_r \phi)^2 + \frac{m^2}{4\pi r^2} (1- \cos \phi) \right)
+ \frac{r^2f(r)}{2e^2} \left( \frac{ e^4 B^2}{16\pi^4} (\phi - \phi_0)^2 + B^2\right)
\nn\ee
The phenomena described in Section \ref{bhsec} are most clearly observed when $r_s \ll r_h$. In this case, the envelope of $\phi$ is well approximated by the exponential profile \eqn{exprofile}. From this, we estimate 
\be \frac{e^2 }{4\pi^2} (\phi-\phi_0) \sim \sqrt{\frac{e^2}{8\pi^2 B}}\partial_{r}\phi \sim \frac{e^2}{4\pi^2} \phi_0 \exp(-r/r_s)
\nn\ee
The kinetic and electric terms in $T_{tt}$ are therefore negligible compared to the magnetic term if
\be
e^2 \phi_0 \ll 1
\nn\ee
We must also consider the energy contained within the kinks.  Since the kinks are embedded in an approximately exponential profile, we estimate that the $n^{\rm{th}}$ kink is located at position $r_{\rm{kink}}$ given by
\be
\phi_0 - 2\pi n \sim \phi_0 \exp(-r_{\rm{kink}}/r_s)
\nn\ee
Demanding that the energy of a kink is small compared to the energy of the magnetic field, we obtain the condition
\be
m_RL e^2 \phi_0 \ll 1
\nn\ee

\section{Appendix: Boundary Conditions and  Filling Fractions}\label{appbound}

In this appendix, we make a few comments about the boundary conditions after bosonization. We start in the UV, where we imposed $\phi=0$. However, it is simple to see that, up to a up to a field redefinition under the shift symmetry $\phi \to \phi + 2\pi$, any regular solution of the differential equation \eqn{did1} must obey this condition automatically.  Asymptotically, as $r\rightarrow 0$, $\phi$ behaves as
\be
\phi \sim Cr^{\Delta_-} + Dr^{\Delta_+} + \frac {\phi_0 r^2}{ (m_R^2L^2 - 2)r_s^2}
\nn\ee
with
\be
\Delta_\pm = \tfrac 1 2 \left(1\pm \sqrt{1+4m_R^2L^2}\right)
\ee
The requirement  that $\Delta_\pm$ is real gives rise to the Breitenlohner-Freedman bound  $m_R^2 L^2> -1/4$ for the bosonized scalar.   Presumably after bosonization, $m_R^2$ is positive. However, it is tempting to speculate that bosonizing the fermions with the standard boundary condition \eqn{fermibc} with $-1/2<m<0$ maps to a negative $m_R^2$ above the BF-bound. If $-\frac 1 4 < m^2_RL^2 < 0$, then both fall-offs of $\phi$ are consistent with the $\phi = 0$ boundary condition, but we can specify a unique solution by demanding that the leading fall-off of $\phi$ is zero.

\subsubsection*{Fractional Filling in the Hard Wall}

In Section \eqn{hwsec}, we imposed the boundary condition $\phi = 2\pi n$, with $n\in {\bf Z}$ at the hard wall. However, as shown in \cite{magcat2}, there is actually a one-parameter family of boundary conditions  allowed at the hard wall. These arise by acting with a chiral symmetry, $\psi_L = -e^{i\theta}\psi_R$ at the boundary.  The effect of this chiral transformation on the boson is simply to shift the boundary  condition to $\phi = 2\pi n + \theta$.  These boundary conditions therefore impose that only fractionally filled Landau levels are allowed.



\section{Appendix: Field configurations in the near-horizon region}\label{appb}

 In Section 3.2, we argued that regularity requires that the scalar field $\phi$ must attain an extremum of the effective potential $V_{IR}(\phi)$ at the black hole horizon.
\be
V_{IR}(\phi) = m_R^2L^2(1-\cos\phi) + \frac{ r_h^2}{r_s^2} (\phi-\phi_0)^2
\nn\ee
When $m^2_RL^2 > r_h^2 / r_s^2$, the effective potential has multiple extrema. To understand this situation further, we shall perform a detailed analysis of the field configurations in the near horizon AdS${}_2 \times \mathbb{R}^2$ region. Converting to the AdS${}_2$ radial coordinate
\be
\zeta = \frac{r_h^2}{6(r_h-r)},
\nn\ee
we obtain the following equation of motion, valid asymptotically near the horizon.
\be
\partial_\zeta^2\phi = \frac{m_R^2L^2}{6\zeta^2} \sin \phi + \frac{r_h^2}{6r_s^2 \zeta^2}(\phi-\phi_0)
\nn\ee
Note that this equation of motion respects the AdS${}_2 \times \mathbb{R}^2$ scaling symmetry $\zeta \to \lambda \zeta$, $t \to \lambda t$, $B \to B$. (The extra factor of 6  comes from the discrepancy between the AdS${}_4$ and AdS${}_2$ radii). Exploiting the scaling symmetry, we cut off the AdS${}_2$ region at $\zeta = 1$ and look for solutions in $\zeta \in [1, \infty)$, satisfying the boundary condition $\phi(\zeta = 1) =0$; other solutions are equivalent to these by scaling.

\para
The  energy of the solution is captured by the effective Hamiltonian, 
\be
H_{AdS_2}=\int_0^\infty d\zeta \left( \partial_\zeta^2 \phi + \frac{m_R^2L^2}{6\zeta^2} (1- \cos\phi) + \frac{r_h^2}{12 r_s^2 \zeta^2}(\phi-\phi_0)^2\right)
\nn\ee
All solutions to this equation of motion tend to a minimum of the effective potential but not
necessarily to the global minimum. The reason is because the red-shift factor ${1}/{\zeta^2}$ means
that one can only gain a finite amount of energy by dropping into a lower minimum of the potential and 
 it is possible for the gradient cost to outweigh the potential gain in making such a change. To illustrate this,  suppose that $\phi_1$ and $\phi_2$ are consecutive minima of $V_{IR}(\phi)$, with $\phi_2 \approx \phi_1 + 2\pi$ and $V_{IR}(\phi_1)> V_{IR}(\phi_2)$. Although $V_{IR}$ is lower at $\phi_2$, it is not necessarily true that a solution that tends to $\phi_2$ at the horizon is lower in energy than a solution tending to $\phi_1$; the field will save electrostatic energy by tending to $\phi_2$, but it costs energy to create the extra kink from $\phi_1$ to $\phi_2$.  Using standard flat space methods to estimate the energy of the kink, one finds that it is energetically favourable to create the extra kink to reach $\phi_2$ if
\be
\frac{8m_RL}{\sqrt{6}} \lesssim \frac{ r_h^2}{12 r_s^2} \left[ (\phi_1 - \phi_0)^2 - (\phi_2 - \phi_0)^2 \right]
\ee
This is related to the pair production bound for fermions in AdS$_2$ \cite{troost}.  If one
ignores the backreaction of the fermions, then it is only viable to place a kink at position $r$ if
the electrostatic gain $\sim {E}/r$ is larger than than the effective mass of the fermion $\sim
{m_R}/r$. Therefore pair production is only viable in $AdS_2$ if one has $E > m_R$.

\begin{figure}[!h]
  \begin{center}
    \includegraphics[trim = 0.5in 0.5in 0.2in 0.7in, width=4.0in]{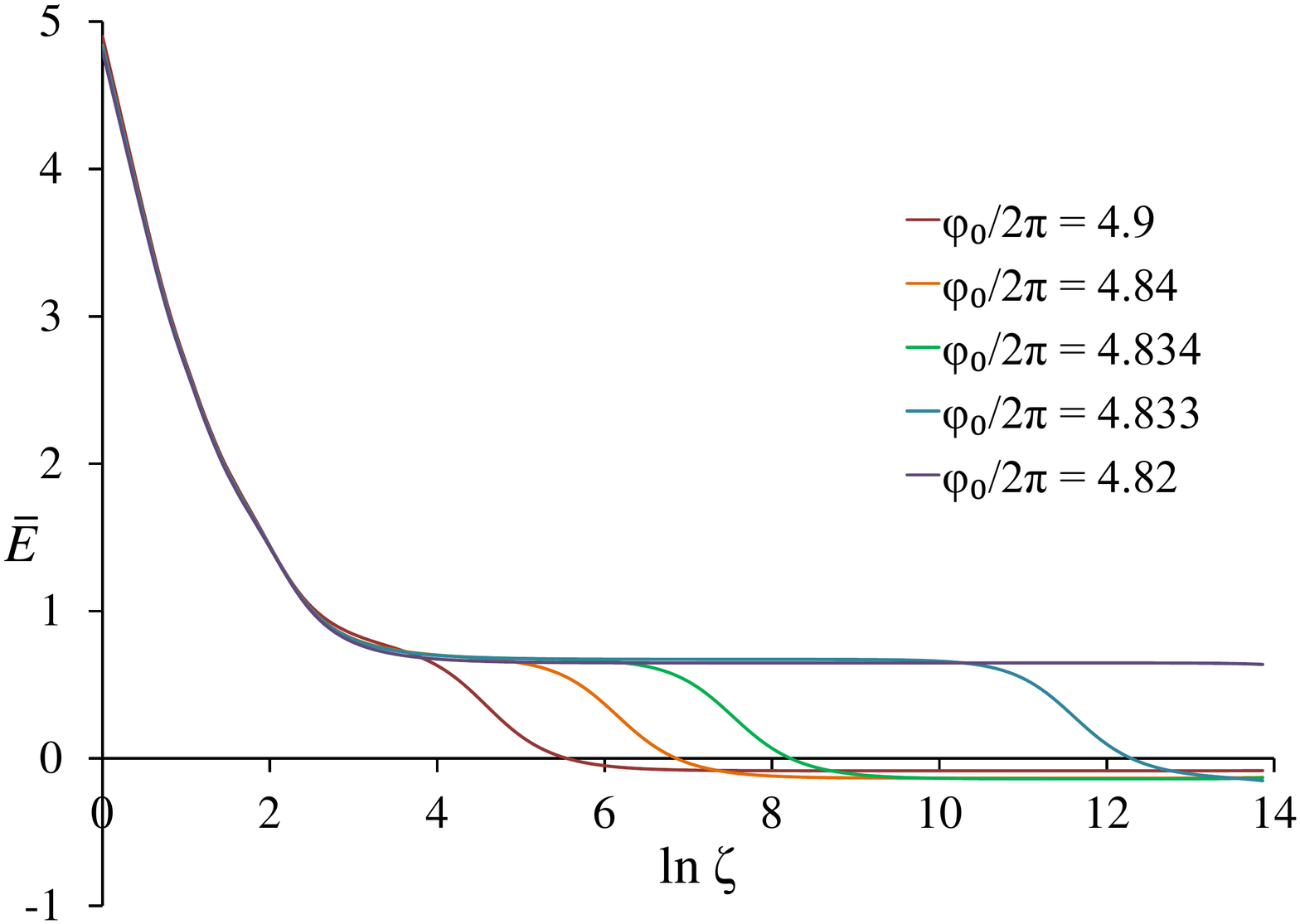}
  \end{center}

  \caption{The electric field, $\bar{E}= r_s^2 F_{rt}/\pi=(\phi_0 - \phi)/{2\pi} $  plotted against logarithm of the AdS${}_2$ coordinate, $\ln \zeta$, for filling fractions ${\phi_0}/{2\pi} = 4.9, 4.84, 4.834, 4.833, 4.82$. The parameters are ${m_R^2}L^2  = 30$ and ${r_h^2}/{r_s^2} = 6$.}
  \label{ads2fig}
\end{figure}

\para
A typical situation is shown in  Figure \ref{ads2fig}. When the filling fraction is in the vicinity of $\phi_0/2\pi \sim 4.83$, $V_{IR}(\phi)$ has two minima at $\phi_1 /2\pi \sim 0.7$, $\phi_2 / 2\pi  \sim -0.1 $, with $V_{IR}(\phi_1) > V_{IR}(\phi_2)$. For $\phi_0 / 2\pi \lesssim 4.833$, the minimum at $\phi_1$ is preferred because the electrostatic energy saved by reaching $\phi_2$ is not enough to compensate for the energy cost of creating the extra kink. Notice also that even though $\phi(\zeta \to \infty)$ changes discontinuously as $\phi_0$ crosses the critical value $4.833$, the solutions $\phi(\zeta; \phi_0)$ vary continuously for $\zeta \in [1,\infty)$ as $\phi_0$ cross this threshold. As $\phi_0$ increases, the position of the kink moves closer and closer to the horizon, and the kink disappears behind the horizon when $\phi_0$ exceeds the threshold.

\noindent


%


\begin{thebibliography}{99}

\bibitem{sungsik}   S.~-S.~Lee,
  ``{\it A Non-Fermi Liquid from a Charged Black Hole: A Critical Fermi Ball},''
  Phys.\ Rev.\ D {\bf 79}, 086006 (2009)
  [arXiv:0809.3402 [hep-th]].
  

\bibitem{mit1}  H.~Liu, J.~McGreevy and D.~Vegh,
  ``{\it Non-Fermi liquids from holography},''
  Phys.\ Rev.\ D {\bf 83}, 065029 (2011)
  [arXiv:0903.2477 [hep-th]].
  
\bibitem{leiden} M.~Cubrovic, J.~Zaanen and K.~Schalm, {\it String Theory, Quantum Phase Transitions and the Emergent Fermi-Liquid},''
  Science {\bf 325}, 439 (2009)
  [arXiv:0904.1993 [hep-th]].
  
 \bibitem{mit2}   T.~Faulkner, H.~Liu, J.~McGreevy and D.~Vegh,
  ``{\it Emergent quantum criticality, Fermi surfaces, and AdS$_2$},''
  Phys.\ Rev.\ D {\bf 83}, 125002 (2011)
  [arXiv:0907.2694 [hep-th]].

\bibitem{review}  N.~Iqbal, H.~Liu and M.~Mezei,
  ``{\it Lectures on holographic non-Fermi liquids and quantum phase transitions},''
  arXiv:1110.3814 [hep-th].


  
\bibitem{hpst}  S.~A.~Hartnoll, J.~Polchinski, E.~Silverstein and D.~Tong,
 ``{\it Towards strange metallic holography},''
  JHEP {\bf 1004}, 120 (2010)
  [arXiv:0912.1061 [hep-th]].


  
\bibitem{sean1} S. Hartnoll, A. Tavanfar, \emph{``Electron stars for holographic metallic criticality,''} Phys. Rev. {\bf D83} 046003 (2011) arXiv:1008.2828v2 [hep-th].

\bibitem{sean2}  S.~A.~Hartnoll, D.~M.~Hofman and A.~Tavanfar,
  ``{\it Holographically smeared Fermi surface: Quantum oscillations and Luttinger count in electron stars},''
  Europhys.\ Lett.\  {\bf 95}, 31002 (2011)
  [arXiv:1011.2502 [hep-th]].

\bibitem{sean3}   S.~A.~Hartnoll, D.~M.~Hofman and D.~Vegh,
  ``{\it Stellar spectroscopy: Fermions and holographic Lifshitz criticality},''
  JHEP {\bf 1108}, 096 (2011)
  [arXiv:1105.3197 [hep-th]].
  
\bibitem{lars1}  V.~G.~M.~Puletti, S.~Nowling, L.~Thorlacius and T.~Zingg,
 ``{\it Holographic metals at finite temperature},''
  JHEP {\bf 1101}, 117 (2011)
  [arXiv:1011.6261 [hep-th]].

\bibitem{lars2}  V.~G.~M.~Puletti, S.~Nowling, L.~Thorlacius and T.~Zingg,
  ``{\it Friedel Oscillations in Holographic Metals},''
  JHEP {\bf 1201}, 073 (2012)
  [arXiv:1110.4601 [hep-th]].
 

  
 

\bibitem{hong} N.~Iqbal, H.~Liu and M.~Mezei,
  ``{\it Semi-local quantum liquids},'
  JHEP {\bf 1204}, 086 (2012)
  [arXiv:1105.4621 [hep-th]].
  
 \bibitem{sachdev} S. Sachdev, \emph{``A Model of a Fermi liquid using gauge-gravity duality,''} Phys. Rev. {\bf D84}, (2011) 066009  arXiv:1107.5321v2 [hep-th].

\bibitem{mcgreevy} A. Allais, J. McGreevy, J. Suh, \emph{``A quantum electron star,''}  arXiv:1202.5308v1 [hep-th].
  
 \bibitem{coleman} S. Coleman, R. Jackiv, L. Susskind, \emph{``Charge Shielding and Quark Confinement in the Massive Schwinger Model''} Annals Phys. textbf{93} (1975) 267. 

\bibitem{adsmono}   S.~Bolognesi and D.~Tong,
  ``{\it Monopoles and Holography},''
  JHEP {\bf 1101}, 153 (2011)
  [arXiv:1010.4178 [hep-th]].

\bibitem{jan1} J.~de Boer, K.~Papadodimas and E.~Verlinde,
 ``{\it Holographic Neutron Stars},''
  JHEP {\bf 1010}, 020 (2010)
  [arXiv:0907.2695 [hep-th]].

\bibitem{jan2} X.~Arsiwalla, J.~de Boer, K.~Papadodimas and E.~Verlinde,
  ``{\it Degenerate Stars and Gravitational Collapse in AdS/CFT},''
  JHEP {\bf 1101}, 144 (2011)
  [arXiv:1010.5784 [hep-th]].


\bibitem{clifford}   T.~Albash, C.~V.~Johnson and S.~MacDonald,
  ``{\it Holography, Fractionalization and Magnetic Fields},''
  arXiv:1207.1677 [hep-th].

\bibitem{albash2}
  T.~Albash and C.~V.~Johnson,
  ``{\it Holographic Aspects of Fermi Liquids in a Background Magnetic Field},''
  J.\ Phys.\ A A {\bf 43}, 345405 (2010)
  [arXiv:0907.5406 [hep-th]].
\bibitem{albash}   T.~Albash and C.~V.~Johnson,
  ``{\it Landau Levels, Magnetic Fields and Holographic Fermi Liquids},''
  J.\ Phys.\ A A {\bf 43}, 345404 (2010)
  [arXiv:1001.3700 [hep-th]].

\bibitem{elena} E.Gubankova, J. Brill, M. Cubrovic, K. Schalm, P. Schijven, J. Zaanen, \emph{``Holographic fermions in external magnetic fields,''}  Phys. Rev. {\bf D84} (2011) 106003  arXiv:1011.4051v2 [hep-th].


 \bibitem{magcat} S. Bolognesi, D. Tong, \emph{``Magnetic Catalysis in AdS${}_{4}$,''} JHEP {\bf 01} (2011) [arXiv:1110.5902v2 [hep-th]].   

\bibitem{magcat2}   S.~Bolognesi, J.~N.~Laia, D.~Tong and K.~Wong,
  ``{\it A Gapless Hard Wall: Magnetic Catalysis in Bulk and Boundary},''
  JHEP {\bf 1207}, 162 (2012)
  [arXiv:1204.6029 [hep-th]].
  
\bibitem{mcreview}
I.~A.~Shovkovy,
  ``{\it Magnetic Catalysis: A Review},''
  arXiv:1207.5081 [hep-ph].

\bibitem{eboli} O. Eboli, \emph{``Abelian Bosonization in Curved Space,''} Phys. Rev. {\bf D36} (1987) 2408.  
    
\bibitem{senechal} D. Senechal, \emph{``An introduction to bosonization,''} (1999) arXiv:cond-mat/9908262 [cond-mat.str-el].

\bibitem{shankar} R. Shankar, \emph{``Bosonization: How to make it work for you in condensed matter,''} Acta Phys. Polon. \emph{B26} (1995) 1835-1867.

\bibitem{urban} F. Urban, A. Zhinitsky, \emph{``Cosmological constant from the ghost: A toy model,''} Phys. Rev. {\bf D80} (2009) 063001, arXiv:1011.2425v2 [astro-ph.CO].

\bibitem{dorn} H. Dorn, \emph{``Path Integral Bosonization of Massive Two-dimensional Fermions''} Phys. Lett. {\bf B167} (1986) 86.



\bibitem{callan2} C. Callan, \emph{``Dyon Fermion Dynamics''}, Phys. Rev. D26 (1982) 2058.




\bibitem{aleksey} A. Cherman and D. Dorigoni, ``{\it Large $N$ and Bosonization in Three Dimensions}" arXiv:1208.1769 


\bibitem{ofer}   O.~Aharony, G.~Gur-Ari and R.~Yacoby,
  ``{\it Correlation Functions of Large N Chern Simons Matter Theories and Bosonization in Three Dimensions},''
  arXiv:1207.4593 [hep-th].

\bibitem{coleman2}  S.~R.~Coleman,
  ``{\it More About the Massive Schwinger Model},''
  Annals Phys.\  {\bf 101}, 239 (1976).

\bibitem{smilga} A. V. Smilga, \emph{``On the fermion condensate in the Schwinger model,''} Phys. Lett. {\bf B278} (1992) 371-376.

\bibitem{dunne}
G.~Basar and G.~V.~Dunne,
  ``{\it The Chiral Magnetic Effect and Axial Anomalies},''
  arXiv:1207.4199 [hep-th].




\bibitem{fraction1} 
  S.~A.~Hartnoll,
  ``{\it Horizons, holography and condensed matter},''
  arXiv:1106.4324 [hep-th].

\bibitem{fraction2} S.Sachdev, ``{\it What can gauge-gravity duality teach us about condensed matter physics?}", 
Annual Review of Condensed Matter Physics 3, 9 (2012), [arXiv:1108.1197]

\bibitem{hartnoll2}  S.~A.~Hartnoll and L.~Huijse,
  ``{\it Fractionalization of holographic Fermi surfaces},''
  arXiv:1111.2606 [hep-th].
 
 \bibitem{koji}  K.~Hashimoto and N.~Iizuka,
  ``{\it A Comment on Holographic Luttinger Theorem},''
  JHEP {\bf 1207}, 064 (2012)
  [arXiv:1203.5388 [hep-th]].





\bibitem{bergman1}  O.~Bergman, N.~Jokela, G.~Lifschytz and M.~Lippert,
  ``{\it Quantum Hall Effect in a Holographic Model},''
  JHEP {\bf 1010}, 063 (2010)
  [arXiv:1003.4965 [hep-th]].

\bibitem{bergman2}  N.~Jokela, M.~Jarvinen and M.~Lippert,
  ``{\it A holographic quantum Hall model at integer filling},''
  JHEP {\bf 1105}, 101 (2011)
  [arXiv:1101.3329 [hep-th]].




 
\bibitem{koslif1}    F.~Denef, S.~A.~Hartnoll and S.~Sachdev,
  ``{\it Quantum oscillations and black hole ringing},''
  Phys.\ Rev.\ D {\bf 80}, 126016 (2009)
  [arXiv:0908.1788 [hep-th]].
  
\bibitem{koslif2} S.~A.~Hartnoll and D.~M.~Hofman,
  ``{\it Generalized Lifshitz-Kosevich scaling at quantum criticality from the holographic correspondence},''
  Phys.\ Rev.\ B {\bf 81}, 155125 (2010)
  [arXiv:0912.0008 [cond-mat.str-el]].



\bibitem{freedman} D. Freedman, A. van Proeyen, \emph{``Supergravity''}, Cambridge University Press (2012).

\bibitem{troost}  B.~Pioline and J.~Troost,
  ``{\it Schwinger pair production in AdS(2)},''
  JHEP {\bf 0503}, 043 (2005)
  [hep-th/0501169].





\end{thebibliography}
\end{document}